\documentclass{article}

\usepackage{arxiv}

\usepackage[utf8]{inputenc} 
\usepackage[T1]{fontenc}    
\usepackage{hyperref}       
\usepackage{url}            
\usepackage{booktabs}       
\usepackage{amsfonts}       
\usepackage{nicefrac}       
\usepackage{microtype}      
\usepackage{lipsum}
\usepackage{amsmath}
\usepackage{subfig}
\usepackage{graphicx}
\usepackage{float}
\usepackage{changepage}
\usepackage{courier}
\usepackage{enumitem}
\usepackage{longtable}
\usepackage{listings}
\usepackage{tikz}
\usetikzlibrary{arrows,shapes,positioning,shadows,trees}
\tikzset{
	basic/.style = {draw, text width=4cm, drop shadow, font=\sffamily, rectangle},
	root/.style = {basic, rounded corners=2pt, thin, align=center, fill=black!20},
	level 2/.style = {basic, rounded corners=6pt, thin, align=center, fill=blue!20, text width=6.5cm},
	level 3/.style = {basic, thin, align=left, fill=pink!40, text width=5.5cm}
}

\def\bs{\boldsymbol}

\newcommand{\pkg}[1]{{\normalfont\fontseries{b}\selectfont #1}}
\let\proglang=\textsf
\let\code=\texttt

\title{\textbf{LISA}: a MATLAB package for Longitudinal Image Sequence Analysis}

\author{Jang Ik Cho \\
			Eli Lilly and Company \\ 
			Global Headquarters Lilly Corporate Center \\ 
			Indianapolis, Indiana 46285 USA\\ 
			E-mail: cho.jangik@gmail.com \\~\\
	\And
	\textbf{Xiaofeng Wang} \\ 
			  Department of Quantitative Health Sciences\\
			  Cleveland Clinic Lerner Research Institute\\
			  9500 Euclid Ave. JJN3\\
			  Cleveland, OH 44139\\
			  E-mail: wangx6@ccf.org	
	\And
	\textbf{Yifan Xu} \\ 
			   Amazon.com, Inc. \\ 
			   410 Terry Avenue N \\
			   Seattle WA 98109 \\
			   E-mail: ethan.yifanxu@gmail.com \\~\\
	\And
	\textbf{Jiayang Sun} \\ 
			   Department of Epidemiology and Biostatistics\\
			   Case Western Reserve University \\ 
			   10900 Euclid Ave. SOM, WG-43\\
			   Cleveland, OH 44106-4945\\
			   E-mail: jsun@case.edu\\
}

\begin{document}
	\maketitle
	
	\begin{abstract}
		Large sequences of images (or movies) can now be obtained on an unprecedented scale, which pose fundamental challenges to the existing image analysis techniques. The challenges include heterogeneity, (automatic) alignment, multiple comparisons, potential artifacts,  and hidden noises.   This paper introduces our \proglang{MATLAB} package, \textbf{L}ongitudinal \textbf{I}mage \textbf{S}equence \textbf{A}nalysis (\pkg{LISA}), as  a one-stop ensemble of image processing and analysis tool  for comparing a general class of images from either different times, sessions, or subjects.   Given two contrasting sequences of images, the image processing in \pkg{LISA} starts with selecting a region of interest in two representative images, followed by automatic or manual segmentation and registration. Automatic segmentation de-noises an image using a mixture of Gaussian distributions of the pixel intensity values, while manual segmentation applies a user-chosen intensity cut-off value to filter out noises. Automatic registration aligns the contrasting images based on a mid-line regression whereas manual registration lines up the images along a reference line formed by two user-selected points. The processed images are then rendered for simultaneous statistical comparisons to generate D, S, T, and P-maps. The D map represents a curated difference of contrasting images, the S map is the non-parametrically smoothed differences, the T map presents the variance-adjusted, smoothed differences, and the P-map provides multiplicity-controlled p-values. These maps reveal the regions with significant differences due to either longitudinal, subject-specific, or treatment changes. A user can skip the image processing step to dive directly into the statistical analysis step if the images have already been processed. Hence, \pkg{LISA} offers the flexibility in applying other image pre-processing tools. \pkg{LISA} also has a parallel computing option for high definition (HD) images.
	\end{abstract}

\keywords{longitudinal images \and high-dimensional data \and image sequence \and multiple comparison \and non-parametric smoothing \and \proglang{MATLAB}}

	\section[Introduction]{Introduction}
	Rapid development in imaging technologies has made medical imaging one of the most important techniques in providing essential evidence for clinical diagnosis and medical intervention. We can now collect unprecedented amounts of image data in both spatial and temporal dimensions per patient, from one or more modalities. The challenges in extracting valuable information from these images motivate modern updates and developments in image processing and analysis methods and tools. For example,  three recent imaging books by Russ and Neal\cite{JRuss2016}, Birkfellner\cite{WBirkfellner2014}, and Burger and Burge\cite{WBurger2016} show the fundamentals of image processing methods developed in the last score. Industries have also invested in developing image processing tools to accommodate many different modalities with types of images ranging from document imaging \cite{docimage} and multimedia imaging \cite{videopross} to medical imaging \cite{leadtools}. The National Institute of Health (NIH)'s Center for Information Technology is leading the {\it Medical Image Processing, Analysis, and Visualization} (MIPAV) project for quantitative analysis of medical images \cite{nihmipav}. Most of the functionalities in MIPAV are for image (pre)processing of  PET, MRI, CT, microscopy, 3D images and other widely-used images in medicine. One of the most widely used image analysis that MIPIV provides, uses summary statistics of the volume of interest (VOI) within a single image to generate contour VOI plots for all of the images to be analyzed.		
	
		Comparing images or assessing their similarities is one of the most important objectives in image analysis. The methods for such comparisons include the methods for pixel by pixel comparison and those based on a specific image measure \cite{Katukam2015}. The measure-based methods include Least Square Image Matching (LSM) \cite{Forstner1982}, Hausdorff distance  \cite{Huttenlochter1993} and components of  a matrix decomposition \cite{Hocenski2000}. Among these measures, LSM has been used extensively because of its simplicity, but it may suffer from inaccurate initial input value and outliers \cite{Forstner1982}.  To obtain comprehensive details of the differences over entire imaging region,  a statistical pixel by pixel comparison is needed.
				
		Denote an  image sequence as spatial and temporal data  $\bs{Y}(s,t,n)$, where $\bs{Y}$ is the intensity value at the pixel or spatial location $\bs{s} \in \bs{S}$ and time $\bs{t} \in \bs{T}$  for the subject $\bs{n} \in \bs{N}$. Here  $\#\bs{S}$, $\#\bs{T}$, $\#\bs{N}$ are the sizes of sets $\bs{S}$, $\bs{T}$, and $\bs{N}$ respectively. Then, the number of measurements  per subject, or the data dimension is $\bs{p} = \#\bs{S} \times \#\bs{T}$. This $\bs{p}$ can be huge, leading to a large-p data problem.  If  $\#\bs{N}$ is much smaller than $p$, we have a large-p-small-n problem. Large-p data challenges statistical multiple comparisons, no matter the number of subjects $\#\bs{N}$ is large or small. How to obtain an objective threshold for multiple comparisons of large, noisy and potentially heterogeneous images has challenged statisticians, especially for analyzing overflowing simultaneous images. Statistical parametric mapping (SPM) \cite{Worsley2002} is an effective and widely-used pixel by pixel image comparison approach.   The motivation for developing  \textbf{L}ongitudinal \textbf{I}mage \textbf{S}equence \textbf{A}nalysis (\pkg{LISA})  MATLAB package comes from generalizing and making a one-stop ensemble of image processing and analysis tool for complex, longitudinal images,  by transforming the components of the Longitudinal Analysis and Self-Registration (LASR) procedure \cite{XWang2006}.  The LASR was developed to overcome the heterogeneity, and the large-p-small-n issue in assessing the effect of Neuromuscular Electrical Stimulation (NMES) experiments \cite{NBergstrom2004} in reducing the risk of developing pressure ulcers.  The data for LASR were movies composed from sequences of seated pressure mat images obtained under a static and a dynamic stimulation, respectively,  in each session from each visit by a patient, for eight patients observed up to 5 years. There were artifacts, heterogeneity, noises, and substantial spatial and temporal alignment issues from different sessions even on the same patient. The data was massive with a huge $p$ and a small $n$ of 8.   These data required extensive imaging preprocessing and general statistical analysis method, Statistical Non-parametric Mapping (SNM) that is more general than the SPM, for each set of comparisons.   Various code for the components of LASR was developed individually and specifically to the pressure data formate. It would have been easier if the \pkg{LISA},  a general purpose MATLAB package was available then.
With increasing amounts of imaging data from different modalities in nowadays, it becomes imperative to implement and generalize the LASR method into a stand-alone MATLAB package for general application of images in standard image formate.  Making \pkg{LISA} available to public as a one-stop MATLAB package sets a basis for future development  for even large data and more generalized case. See the discussion in Section 6.  

\begin{figure}[htb]
		\centering
		\includegraphics[width=\textwidth]{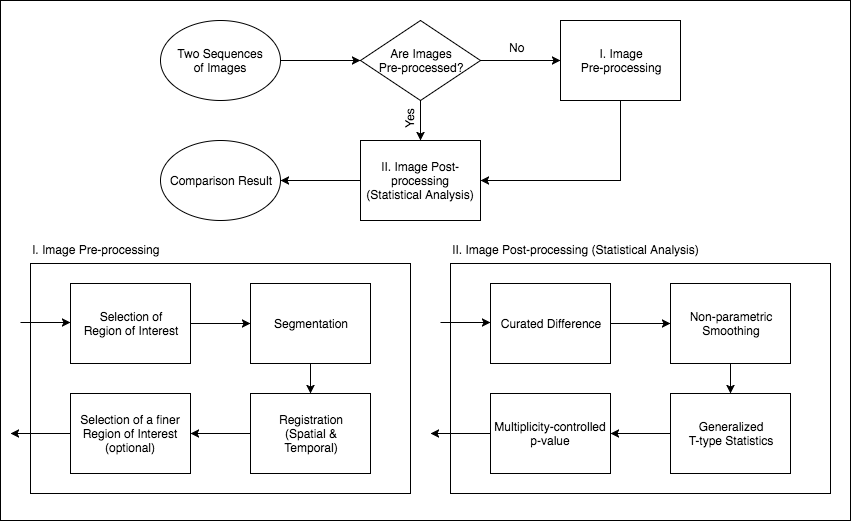}
		\caption{Structure of \pkg{LISA}}
		\label{fig:structure}
	\end{figure}

         Figure \ref{fig:structure} above shows the schematic  structure of \pkg{LISA} tool, where the `image processing' is called `image pre-processing,' and `statistical analysis' is called `image post-processing.' The top part of Figure \ref{fig:structure} presents the overall flow of \pkg{LISA}. A complete process of \pkg{LISA} goes through 2 steps: ``I. Image Pre-processing'' step as introduced in the left bottom flowchart of Figure \ref{fig:structure} and ``II. Image Post-processing (Statistical Analysis)'' step as introduced in right bottom flowchart of Figure \ref{fig:structure}. However, if image pre-processing is unnecessary or a user chooses to use other \proglang{MATLAB} built-in tools to de-noise, segment or register their two sequences of images, \pkg{LISA} has the option to skip the image pre-processing step to go directly to image post-processing (statistical analysis). Image pre-processing is done by 1) selecting a region of interest (ROI), 2) de-noising the images by automatic or manual image segmentation, and 3) aligning the images by also automatic or manual image registration. Image post-processing is done automatically by 1) calculating curated differences, 2) de-noising the differences obtained from Step 1 by a non-parametric smoothing, 3) deriving variance-adjusted, smoothed values to get T-type statistics, and 4) providing final significances of the differences based on the multiplicity adjusted p-values. These four post-processing  steps provide Statistical Nonparametric Mapping (SNM) with minimal assumptions on the images. Here we use the False Discovery Rate (FDR) to adjust the multiplicity. 

 The development of \proglang{MATLAB} package \pkg{LISA}(Longitudinal Image Sequence Analysis) has implemented and  expanded the functionality of LASR to general cases for any images that can be aligned using a reference line (estimable from the image data directly)  or general image processing tools in Matlab. A graphical user interface in \pkg{LISA}  provides convenience for a user to quickly remove noise and artifacts, decide on the region of interests and choose a quick reference line if feasible. In addition, \pkg{LISA} provides a parallel computing option for a large sequence or high-resolution images.  

	In the remaining paper, we provide the \pkg{LISA} installation guide  (\S2),   options for pre-processing (\S3) and  post-processing  (statistical analysis)  with  a brief introduction of methodologies (\S4),  demonstrate the  software \pkg{LISA}  via an application (\S5), and then conclude the article with a discussion, including limitations and further extensions (\S6).   For the details and interpretations of  of  the D, S, T and P maps, see Section 4.  With features introduced in  \pkg{LISA}, we improve the accuracy of statistical comparison for both static and dynamic sequences of images. \pkg{LISA} can tackle two sequences of time-series images in general, including seated pressure mat images used in the NMES study, time-series satellite image\cite{JVerbesselt2010}, and time-series chemical imaging data \cite{Gowen2011}.

	\section[Installation]{Installation}
	A step-by-step installation guide is provided in this section.\\
	
	\textbf{1. Download \pkg{LISA} App} \\
	
	\pkg{LISA} program ``LISA.zip'' can be downloaded from the following two locations.
	
	\begin{description}
		\item[$\cdot$] MathWorks File Exchange website : \url{http://www.mathworks.com/matlabcentral/fileexchange/?term=type%3AApp}
		\item[$\cdot$] Center for Statistical Research, Computing, and Collaboration(SR2c) website : \url{http://sr2c.case.edu/software.html}
	\end{description}

	The archive ``LISA.zip'' contains three files : LISAxxx.mlappinstall, LISAxxx.prj, and ReadMe where xxx is the package version number. ``LISAxxx.mlappinstall'' file is the installation file, ``LISAxxx.prj"' is a project file containing package information and ``ReadMe'' file is a quick starting guide.\\

	\textbf{2. Install and load \pkg{LISA} App} \\
	
	With a newer version of \proglang{MATLAB} (version 8 or above), one can install and load \pkg{LISA} by simply running the LISAxxx.mlappinstall file \cite{installpackage} and follow the three steps below. However, with an older version of \proglang{MATLAB} (version 7 or below) one can manually install \pkg{LISA} using the  instructions provided in the following MathWorks website \url{https://www.mathworks.com/help/matlab/matlab_env/manage-your-add-ons.html}.
	
\begin{figure}[H]	
	\begin{minipage}{.5\linewidth}
		\centering
		\subfloat[][Installation Window]{\includegraphics[width=4.5cm]{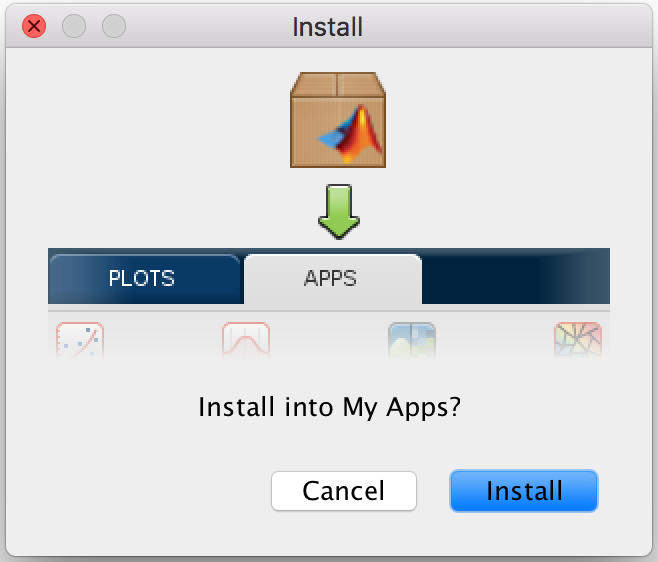}\label{<figure1>}}
	\end{minipage}
	\begin{minipage}{.5\linewidth}
		\subfloat[][Installation Complete]{\includegraphics[width=4cm]{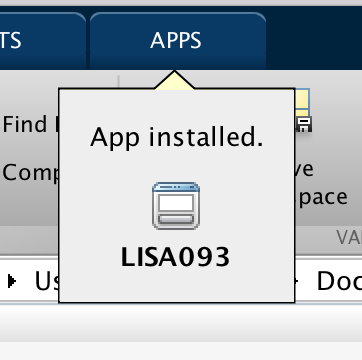}\label{<figure2>}}
	\end{minipage}
		\caption{Installation Process}
		\label{fig:install}
	\end{figure}

\begin{figure}[H]
		\centering
		\includegraphics[width=10cm]{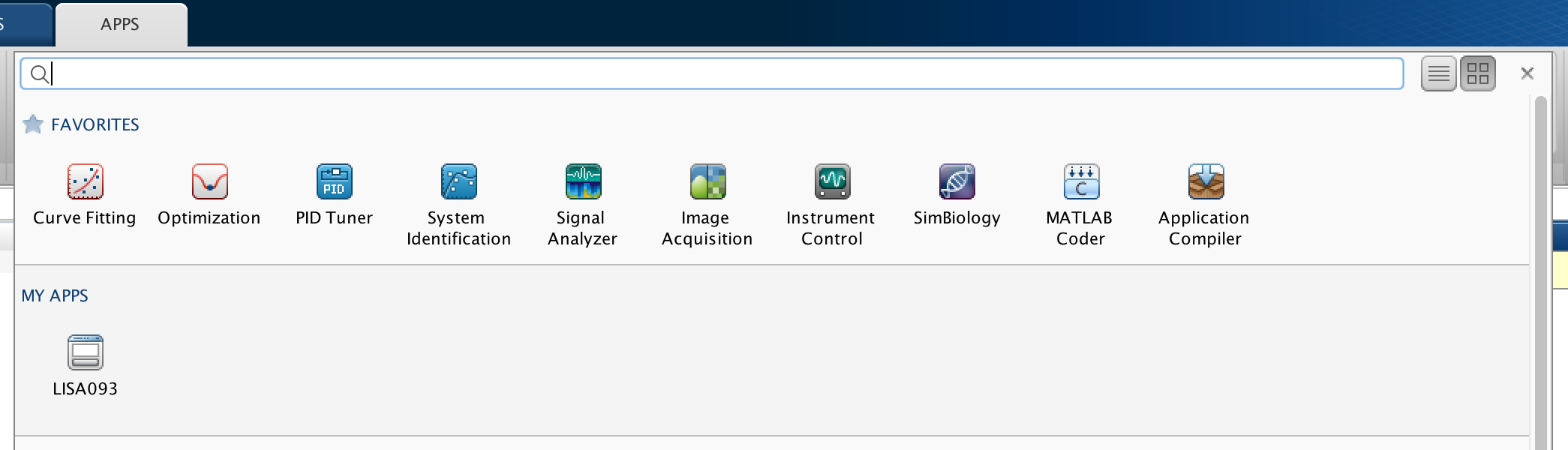}
		\caption{Where \pkg{LISA} can be found and launched}
		\label{fig:package}
	\end{figure}

	\begin{enumerate}
		\item Run LISAxxx.mlappinstall file. Then \proglang{MATLAB} will be launched and a ``Install'' window will pop-up as shown in Figure \ref{fig:install} (a)
		\item Continue by clicking the ``Install'' button. \proglang{MATLAB} will automatically install the package and will notify the installer from the ``APP'' tab when the installation is complete, as shown in Figure \ref{fig:install} (b)
		\item Load \pkg{LISA} by clicking on the \pkg{LISA} App icon under ``APPS - MY APPS'' as shown in Figure \ref{fig:package}
	\end{enumerate}

	When \pkg{LISA} is loaded, a user can start using the program right away by running the \code{lisa} function. Typing ``helpLISA'' into \proglang{MATLAB} command window will provide a vignette including the syntax and examples of the \code{lisa} function.	
	
\section[Dataprocessing]{Data and Image Processing}
\pkg{LISA} is bulit to compare two sequences of images by contrasting each sequence with representing images taken in one session. The first stable image of each sequence is treated as its reference image, which will be used throughout the image processing. Data and image processing consists of the following 5 steps.
\begin{enumerate}[nolistsep]
	\item Data Scan : Scan image frames from ``csv" files
	\item Region of Interest(ROI) Selection : Crop images to the region of interest
	\item Image Segmentation : Detect noise
	\item Image Registration : Calibrate Images
	\item Finer ROI Selection(optional)
\end{enumerate}

\subsection{Data Scanning}
\begin{figure}[H]	
	\begin{minipage}{.325\linewidth}
		\centering
		\subfloat[Row Identifier]{\includegraphics[width=5cm]{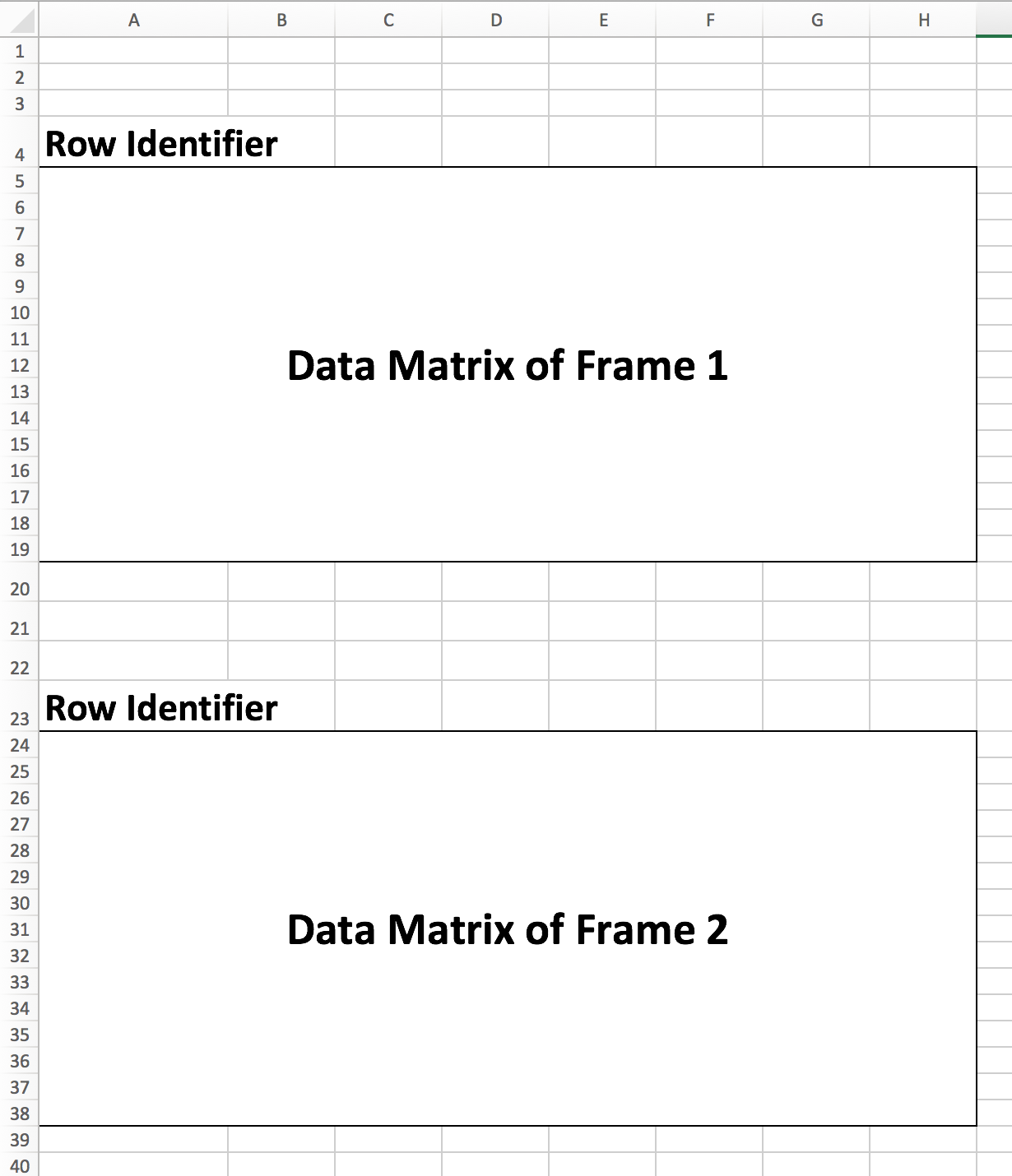}\label{<figure1>}}
	\end{minipage}
	\begin{minipage}{.325\linewidth}
		\centering
		\subfloat[Column Identifier]{\includegraphics[width=5cm]{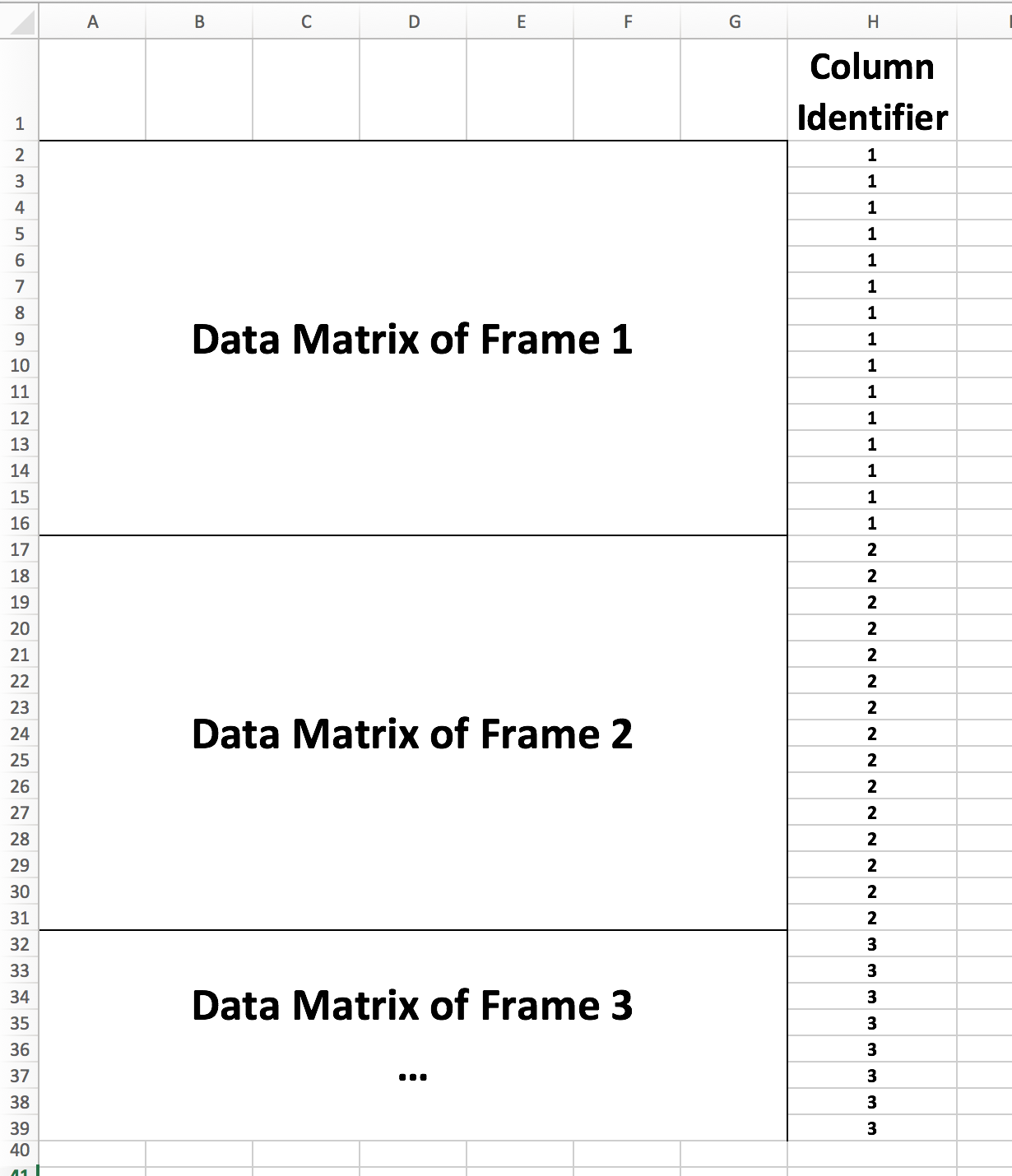}\label{<figure2>}}
	\end{minipage}
	\begin{minipage}{.325\linewidth}
		\centering
		\subfloat[Blank Row]{\includegraphics[width=5cm]{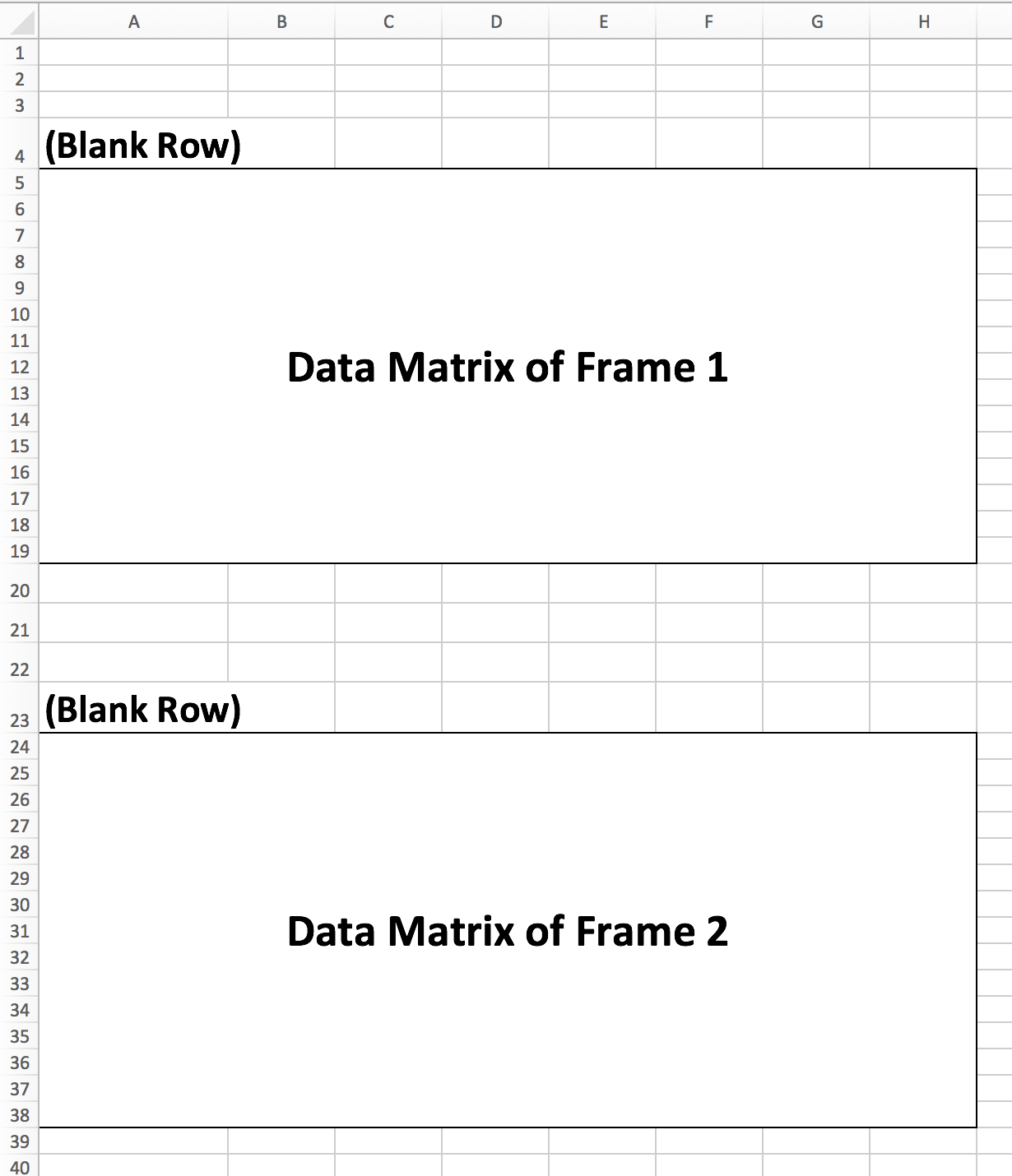}\label{<figure3>}}
	\end{minipage}
	\caption{Data Matrix Scanning Method}
	\label{fig:scandata}
\end{figure}

The current version of \pkg{LISA} accepts ``csv" files as inputs. A user can convert various image formats to a ``csv'' format in \proglang{MATLAB} by reading images using \code{imread} and saving the image to a ``csv'' file using \code{csvwrite}. In general, the output of an imaging tool in a ``csv'' file includes information on the date, time, frame numbers, intensity values of the images and metadata such as equipment model, user information etc. Therefore, it is very important to distinguish the image intensity matrix in a numerical form to conduct statistical comparison. \pkg{LISA} is able to extract image intensity matrices from the raw data files that are in the following three formats as shown in figures \ref{fig:scandata}(a), \ref{fig:scandata}(b), and \ref{fig:scandata}(c). 

The data structure in Figure \ref{fig:scandata}(a) requires a user to extract the data matrix using a row identifier. A user needs to set the parameter \code{`scantype'} to \code{`row'}  and a row identifier character string needs to be passed over to the parameter \code{`rowid'}. \pkg{LISA} skips the header rows until it encounters the first identifier. It then scans next \texttt{`nrow'} number of rows to extract as a numeric matrix and define it as the first image frame. \pkg{LISA} continues the same steps until it reaches the last identifier. Here is an example of the syntax for scanning the data using the row identifier having all other arguments with their default values.\\

\code{lisa(`file1.csv',`file2.csv',`row',\#frame,\#row,\#col,`rowidentifier',\\`',`',`',`',`',`',`',`' )}\\
		
The data structure in Figure \ref{fig:scandata}(b) requires a user to extract the data matrix using a column identifier. A user needs to specify the column that contains information on the frame numbers by setting the parameter \code{`scantype'} to \code{`col'}  and a column number needs to be passed over to the parameter \code{`colid'}. \pkg{LISA} reads the number in that column to extract numeric matrix and save each matrix to its corresponding frames. Here is an example of the syntax for scanning the data using the column identifier having all other arguments with their default values.\\
		
\code{lisa(`file1.csv',`file2.csv',`col',\#frame,\#row,\#col,` ',\# of column\\identifier,`',`',`',`',`',`',`',`' )}\\
		
The data structure in Figure \ref{fig:scandata}(c) requires a user to extract the data matrix using a blank row right before the numeric intensity matrix. A user needs to set the parameter \code{`scantype'} to \code{`blank'} and \pkg{LISA} skips the header rows until it encounters the first blank row. It then scans next \code{`nrow'} number of rows to extract as a numeric matrix and define it as the first image frame. \pkg{LISA} continues the same steps until it reaches the last blank row. Here is an example of the syntax for scanning the data using a blank row having all other arguments with their default values.\\
		
\code{lisa(`file1.csv',`file2.csv',`blank',\#frame,\#row,\#col,`',`',`',`',\\`',`',`',`',`' )}	
		
\subsection{Select Region of Interest}
This is the first interactive step for a user. A user is given the opportunity to select a rectangular region of interest from the first images of each sequence. Cropping images to a specific area of interest enhances the result of data segmentation and data registration because this naturally eliminates some part of the noise. \pkg{LISA} uses the ``\code{imcrop}" function within the \proglang{MATLAB Image Processing Toolbox} to select the region of interest.
		
\subsection{Image Segmentation}
 It is critical to reduce noises and outliers by segmenting an image, which  sets the noisy intensity values to zero before performing an image registration. LISA provides data-driven automatic and manual segmentation. A  pixel is identified as noise if its intensity value is less than a threshold $C$. Let $Y (i, j)$ denote the pixel intensity value at the $i$th row and $j$th column of a representative image. Then the intensity values in the segmented image are  $\widetilde{Y}(i,j)$ below:	
\begin{equation}
	\tilde{Y}(i,j) = \begin{cases}
	Y(i,j) &\textrm{if $Y(i,j)$ > C}\\
	0 &\textrm{otherwise}
	\end{cases}.
\end{equation}
		The  threshold value  $C$  is derived differently for automatic segmentation and manual segmentation, as shown below. 
		
\subsubsection{Automatic Segmentation}
Automatic segmentation finds an optimal $C$ that separates data into two groups with the maximum separation, based on an estimated mixture of Gaussian distributions of  the pixel intensity values, as introduced in \cite{XWang2006}. The mixture density $f(y)$ has the following form:
\begin{equation}
	f(y) = \displaystyle\sum_{g=1}^{G} \frac{\alpha_g}{\sigma_g}\phi\left(\frac{y - \mu_g}{\sigma_g}\right)\equiv\alpha_1\cdot f_1(y)+ (1-\alpha_1) \cdot f_2(y)
\end{equation}
			where $\phi$ is the standard normal density, $G$ is the number of mixtures,  and $\{\alpha_g > 0, \;g=1, ..., G\}$  are mixing parameters satisfying  $\sum_{g=1}^{G} \alpha_g = 1$. The first-component density $f_1(y) = \phi\left[(y - \mu_1)/\sigma_1\right]/\sigma_1$ represents the background distribution, while  the second-component density $f_2(y) = \sum_{g=2}^{G}(\alpha_g/\sigma_g)\cdot \phi\left[(y - \mu_g)/\sigma_g\right]/(1-\alpha_1)$ represents the distribution of pixel intensity inside the region of interest, also as a mixture of Gaussian distributions. The $C$ value automatically computed by LISA separates $f_1$ and $f_2$ at the valley bottom  between the two densities -  see the black vertical line in Figure 5.
			
\begin{figure}[H]
	\centering
	\includegraphics[width=9cm]{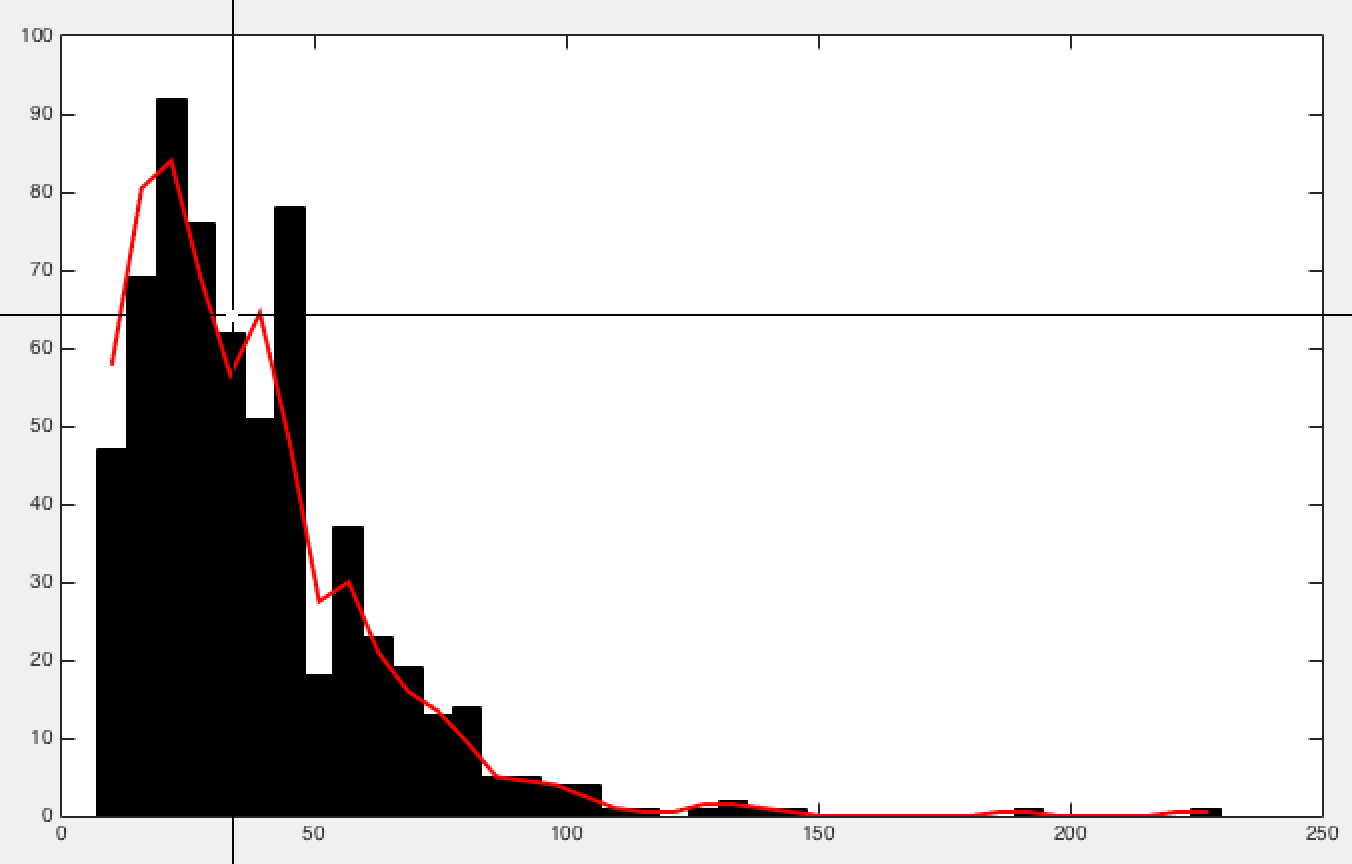}
	\caption{Data Segmentation from Histogram and Density Plot}
	\label{fig:segexampleplot}
\end{figure}
			
Figure \ref{fig:segexampleplot} is an example of the histogram and density plot for the pixel intensity values in a seated pressure map. The distribution of the intensity values is modeled by a mixture of 3-component Gaussian distributions. The threshold $C$ is determined by the vertical line that separates the first and second components in  the fitted parametric Gaussian mixture model.

\subsubsection{Manual Segmentation}
\begin{figure}[H]
	\centering
	\includegraphics[width=9cm]{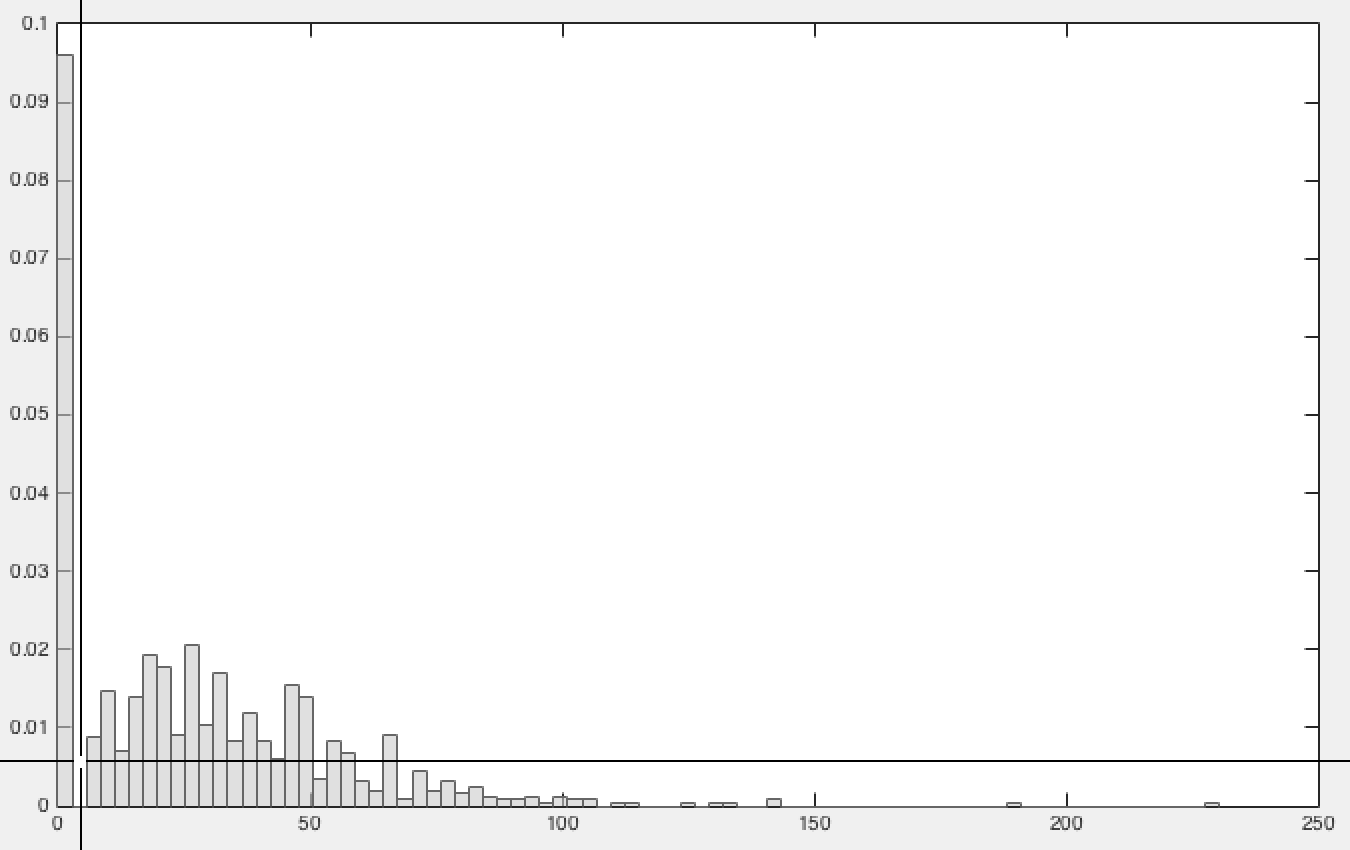}
	\caption{Manual Segmentation from Histogram}
	\label{fig:manuseg}
\end{figure}

Manual segmentation allows a user to choose  $C$ visually based on a histogram of pixel intensity values. Figure \ref{fig:manuseg} is a histogram of intensity values.  It is natural to select the threshold $C$ at the gap where the target cross is located because at that threshold value the intensity values are distinctively divided into two groups. Once the threshold $C$ is selected, intensity values below $C$ will be replaced with zero, defined as the background noise.
			
\subsection{Image Registration}
There are two steps in image registration: temporal registration and spatial registration. Temporal registration excludes unstable images from each sequence and identifies the most similar subset of images to compare between the two sequences. Spatial image registration is done using a geometric transformation of the images. 

\subsubsection{Temporal Registration}
The  correlation coefficient of the intensities between two images is  a reasonable measure of similarity between the two images. \pkg{LISA} temporally aligns two image sequences (of same speed) by shifting one sequence  forward $j$
frames to match temporally the conditions of another sequence, using an Intensity-based Correlation Registration (\textbf{ICR})  below.
\begin{enumerate}
	\item Exclude $5$ or the first 5$\%$ of images from each sequence, whichever the number is larger. This step  removes  noise images  at the start of recording.  Let $S^1_1, ..., S^1_n$ and $S^2_1,...,S^2_n$ be the emaining  $n$ sequential images in  two image sequences $S^1$ and $S^2$.
	\item Compute the shift-$j$ correlation coefficient of two sequences by averaging correlation coefficients of  the  data from the  respective $i$-th and $(i+j)$th frames of two sequences over $i = 1,...,n-j$, for each $j = 0,...,n-1$  to obtain:
		$$AvgCor_j = \dfrac{1}{n-j} \sum_{i} cor(S^1_i,S^2_{i+j})$$
	Then find $j_{max}$ that maximizes the correlations coefficients $vgCor_j$ over $j$. In other words  $j_{max}= \min\{j_0:  AvgCor_{j_0} = \max_{j}(AvgCor_j)\}$.
	\item Align $S^1_i$ and $S^2_{i+j_{max}}$ for  frames $i = 1,...,n-j_{max}$.
\end{enumerate}

\subsubsection{Spatial Registration}
\pkg{LISA} provides automatic registration and manual registration for the sequences after a temporal registration,  
		
\paragraph{Automatic Registration}
The following midline registration algorithm  from Algorithm 3.1 in \textit{Wang et al.(2006)}\cite{XWang2006} is applicable to registering  images that can be aligned by a rigid transformation using a midline (and a point) as a natural landmark.

\setlength{\leftskip}{0.4cm} \textbf{Automatic Registration Algorithm}:
\begin{enumerate}
	\item Determine, column by column, the mid-points($m_c$) by counting non-zero values in each column of an image using $m_c$ = $\frac{r}{2} + \frac{(u - l)}{2}$, where $r,u,l$ are the number of rows, number of non-zero intensity values from the upper half and lower half of the column respectively for $c = 1,...,\# of columns$.
	\item Fit a regression line through the mid-points and call this line the mid-line.
	\item Conduct a rigid transformation by aligning each mid-line with a horizontal line as shown in Figure \ref{fig:autoregt} using a rotation and location shift.
		
	\begin{figure}[H]
		\centering
		\includegraphics[width=9cm]{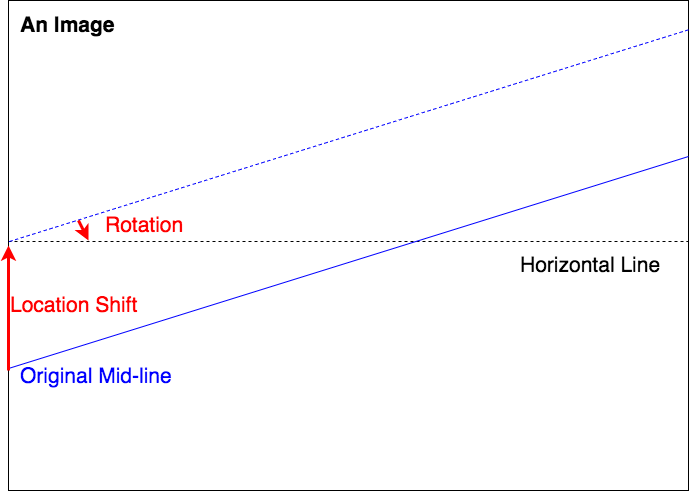}
		\caption{Transformation in Automatic Registration}
		\label{fig:autoregt}
	\end{figure}			
			
	In other words, we define the transformation matrix \textbf{R} using the slope and intercept estimates of the mid-line from step 2 as 
	$$\textbf{R} = \begin{bmatrix} cos\theta&-sin\theta&s_x\\ sin\theta&cos\theta&s_y\\ 0&0&1 \end{bmatrix}$$
	where $tan\theta$ is the slope of the mid-line and $s_x$ and $s_y$ are the shifting parameters. Using $R$ we change $X$ to $X' = RX$ where $X$ contains the information of the coordinate of the pixel defined as $X = [x,y,1]$.	
\end{enumerate}
		
\begin{figure}[H]
	\begin{minipage}{.5\linewidth}
		\centering
		\subfloat[][Before Registration:Image and midline]{\includegraphics[width=6cm]{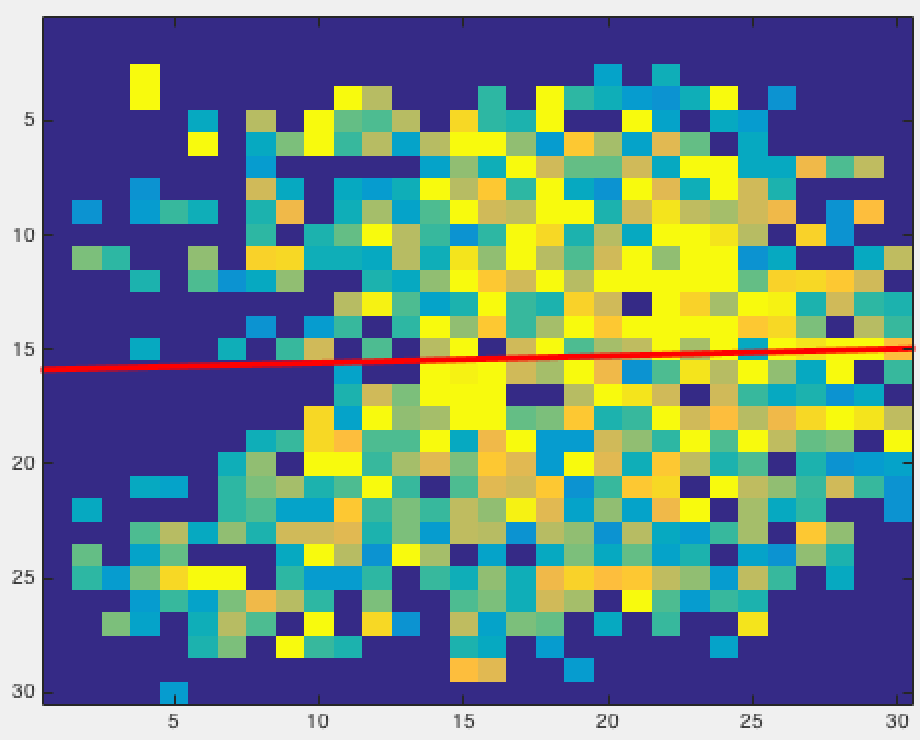}\label{<figure1>}}
	\end{minipage}
	\begin{minipage}{.5\linewidth}
		\subfloat[][After Registration:Image and 	midline]{\includegraphics[width=6cm]{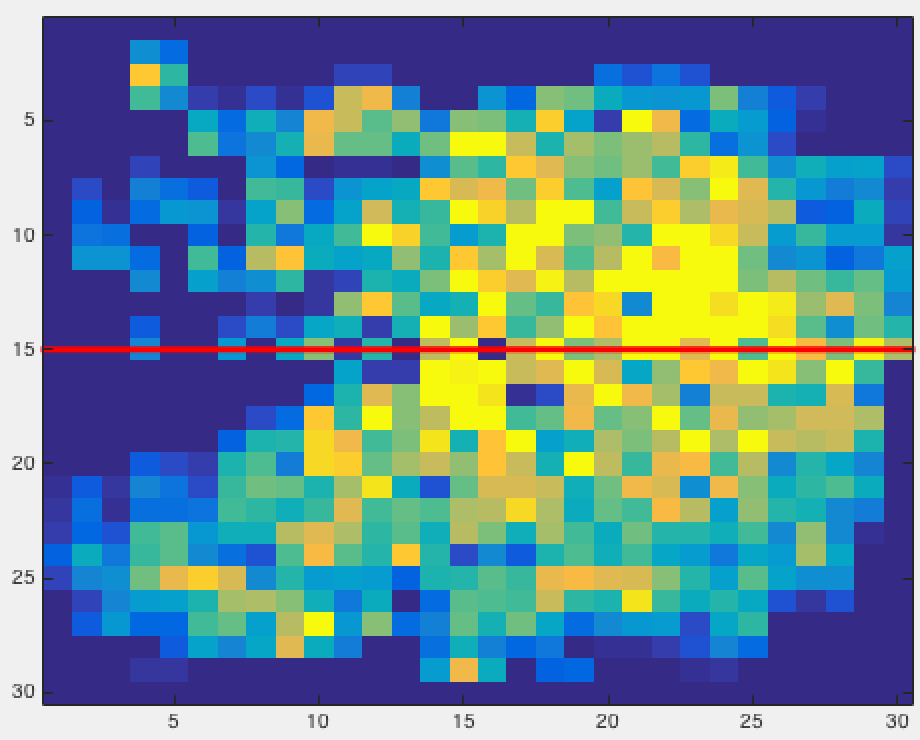}\label{<figure2>}}
	\end{minipage}
	\caption{An example of automatic spatial image registration}
	\label{fig:autoregistplot}
\end{figure}

Figure \ref{fig:autoregistplot} shows an example of automatic registration. The image is translated to be aligned with the horizontal center line. The mid-line registration algorithm given above is fast to implement automatically when comparing sequences of many images. For images that require a rescaling or other types of registration, we suggest a user to use other \proglang{MATLAB} built-in image processing tools or manual registration in the next section.
		
\begin{figure}[H]
	\begin{minipage}{.5\linewidth}
		\centering
		\subfloat[][Points selection for the 1\textsuperscript{st} sequence]{\includegraphics[width=6cm]{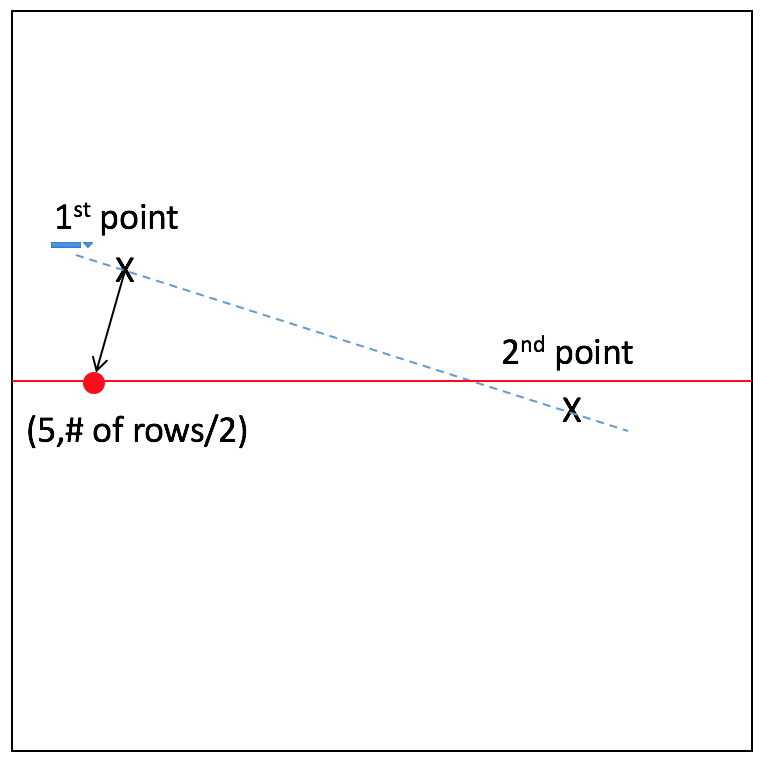}\label{<figure1>}}
	\end{minipage}
	\begin{minipage}{.5\linewidth}
		\subfloat[][Points selection for the 2\textsuperscript{nd} sequence]{\includegraphics[width=6cm]{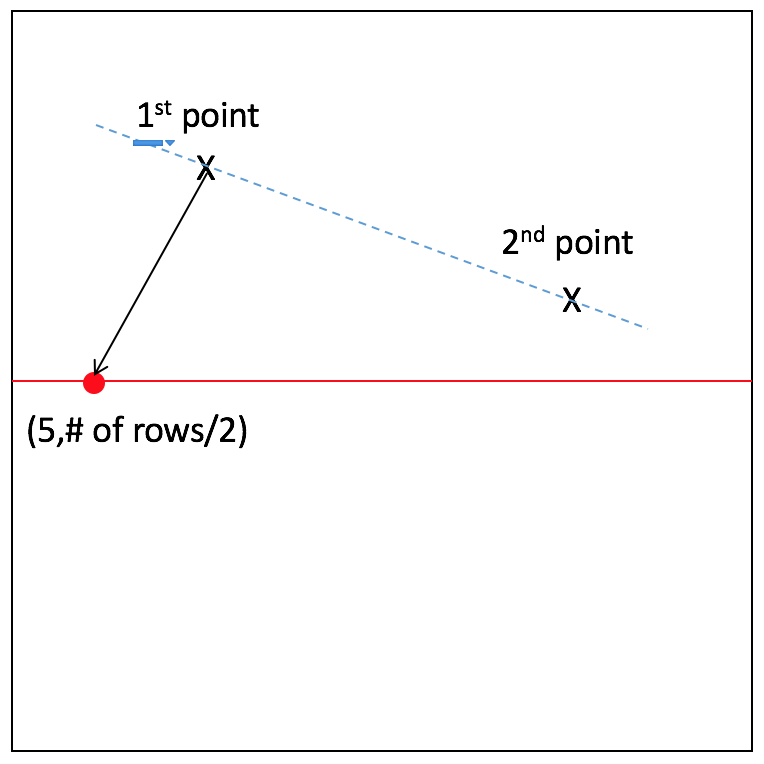}\label{<figure2>}}
	\end{minipage}
	\caption{An example of point selection in manual registration}
	\label{fig:manuregistex}
\end{figure}

\paragraph{Manual Registration}
Manual registration aligns the images using a line defined by two user-selected points. Users select two points on the first images from each sequence which makes four point selections in total. The first points selected for each sequence are used as a reference point which is aligned at (5,\# of rows/2) to shift the image to the same starting point. The second points selected is used to form a line which is used to rotate the image to the horizontal middle line of the image. An illustration of the point selection is provided in Figure \ref{fig:manuregistex}. 
		
Then \pkg{LISA} will perform a rigid transformation based on a shift and rotation matrix of \textbf{R} defined by the dotted blue line and 1\textsuperscript{st} points from Figure \ref{fig:manuregistex}, where \textbf{R} = $\begin{bmatrix} cos\theta&-sin\theta&u\\ sin\theta&cos\theta&v\\ 0&0&1 \end{bmatrix}$. The mathematics behind the transformation using the matrix \textbf{R} us identical to step 3 in the automatic registration. 	
	
\subsection{Optional ROI selection}
\begin{figure}[H]
	\begin{minipage}{.5\linewidth}
		\centering
		\subfloat[][Before optional ROI Selection]{\includegraphics[width=6cm]{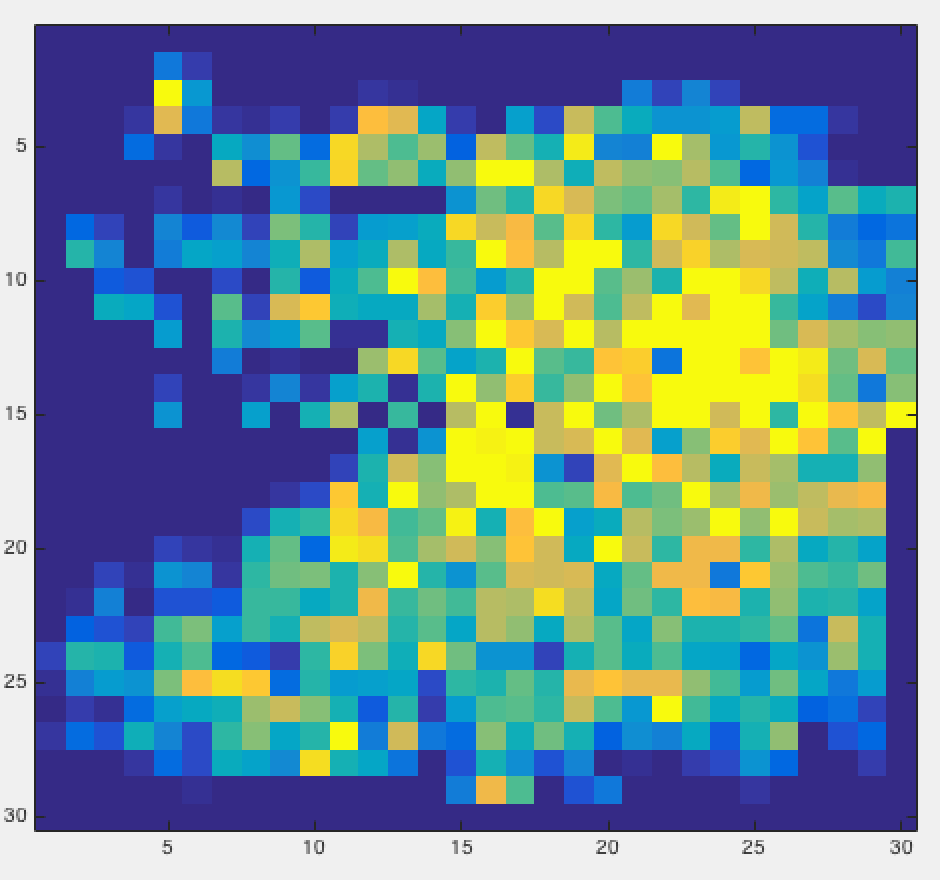}\label{<figure1>}}
	\end{minipage}
	\begin{minipage}{.5\linewidth}
		\subfloat[][After optional ROI Selection]{\includegraphics[width=6cm]{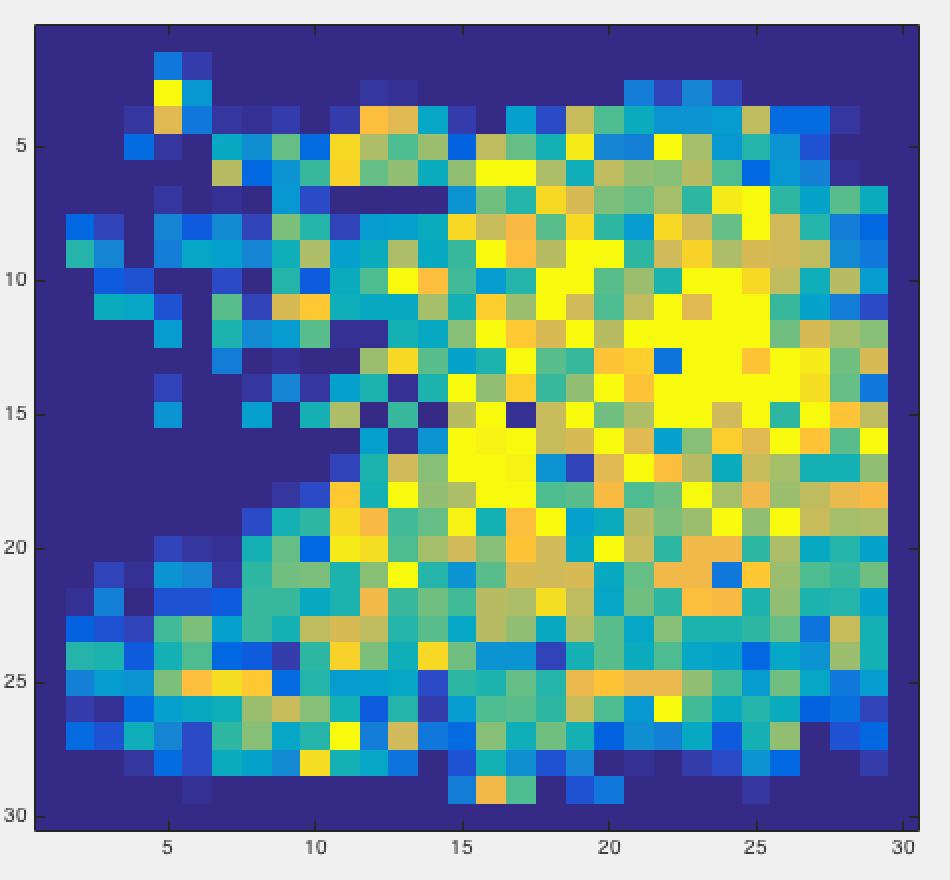}\label{<figure2>}}
	\end{minipage}
	\caption{Example of optional ROI selection.}
	\label{fig:froiresult}
\end{figure}

\pkg{LISA} provides an optional region of interest selection after the image registration as shown in Figure \ref{fig:froiresult}. This step is done by connecting a polygon around the ROI to distinguish ROI from the background by setting intensity values outside of the ROI to zero. Figure \ref{fig:froiresult} shows an example of optional ROI selection after registration. Comparing figure \ref{fig:froiresult}(a) and figure \ref{fig:froiresult}(b), regions at y = 7 to 15 at x = 30 of the image in figure \ref{fig:froiresult}(b) is trimmed off after the selection.
		
This step is optional to be used only when it is necessary because a user needs to be cautious when selecting the ROI at this step. If the ROI is accurately selected, it will improve the accuracy of our statistical analysis. However, if the ROI selection includes artifacts or strong noises by accident, it will corrupt the image and possibly lead to misleading significance regions.
	
\section[Statanalysis]{Statistical Analysis}
Our primary objective for \pkg{LISA} is to identify the areas where there have been significant changes. The statistical methods to achieve this goal are given in the following four steps, by revising the Algorithm 4.1 in our early work \cite{XWang2006}.
	
\begin{enumerate}[nolistsep]
	\item {Curated difference}
	\item {Non-parametric smoothing map}
	\item {Generalized T-Map}
	\item {FDR-controlled P-Map}
\end{enumerate}
	
\subsection{Curated Difference}
The initial step is to calculate curated difference of intensity values between the pairs of images in the two sequences being compared. For an image with a resolution of $n$ by $m$, the image contains $N$ number of pixels where $N$ = $n \times m$. Let $Y_1(i,j,t)$ and $Y_2(i,j,t)$ be the pixel intensity values at coordinate $(i,j)$ of the corresponding images at time $t$ in sequence 1 and sequence 2 respectively. Then the curated difference $d(i,j,t)$ is a pixel by pixel subtraction defined as $d(i,j,t) = Y(i,j,t) - Y(i,j,t)$ for $i = 1, ..., n$, $j = 1, ..., m$ and $\forall{t}$ are calculated for every pair of image comparison from the two image sequences and are used to generate the time series D-maps.
		
\subsection{Non-parametric smoothing map}
Our non-parametric smoothing map first pads the border of curated difference and then smooth the curated difference, $d_{j}$,  in  the region of interest. Padding the edge will prevent outstanding false-positive differences due to image alignment and noise. Our non-parametric smoothing method uses a local polynomial regression. The method is reproduced from 4.1.T-type Statistics \cite{XWang2006} as follows. 
		
For a single set of curated difference image data $\{ (x_{1j}, x_{2j}, d_j); j=1,...,N\}$, consider  the model
$$d_{j} = m(x_{1j}, x_{2j}) + \epsilon_j$$
where $m(\cdot)$ is an unknown smooth function and $\epsilon_i$ is an error term representing the random errors of the curated differences derived from the image observations. Then the  smooth function $m$ at
$s=(x_1, x_2)$ can be approximated by the value of  a  fitted local polynomial function using the intensity values in the  neighborhood of $s$. Specifically, for any $s=(x_{1}, x_{2})$ of interest, the least squares problem seeks  $\hat{\bs{\beta} }= (\hat\beta_0, ..., \hat\beta_5)$ such that
\begin{eqnarray*}
\hat{\bs\beta} &=&\min^{-1}_{\bs{\beta}} \sum_{j=1}^N \left\{d_j - \beta_0 - \beta_1(x_{j1}-x_1) - \beta_2(x_{j2}-x_2) - \frac{\beta_3}{2}(x_{j1}-x_1)^2 - \frac{\beta_4}{2}(x_{j2}-x_2)^2\right.\\
& &\left.\qquad  \qquad \qquad- \beta_5(x_{j1}-x_1)(x_{j2}-x_2) \right\}^2 w(x_{j1}-x_1; h_1) w(x_{j2}-x_2; h_2)
\end{eqnarray*}
and then defines the estimate of the regression surface as $\hat{m}(x_1,x_2) = \hat{\beta}_0$ \cite{WCleveland1988,XWangNon2010}. The smooth weight function $w(z; h)$ peaks at $z=0$ and the smoothness of $\hat{m}$ is controlled by the bandwidth $h$. A normal density function with mean 0 and standard deviation $h$ is a typical choice. In our procedure, the bandwidths $h_1$ and $h_2$ are selected based on the approach of 10 fold cross validation with 20 iterations \cite{Hastie2001}.
		
\subsection{Generalized T-Map}		
The method used for generalized T-Map is reproduced from 4.1.T-type Statistics \cite{XWang2006} as follows. \\
The hypothesis we are testing is $H_0 : m(s) = 0$ vs. $H_1 : m(s) > 0$ at $s = (x_1, x_2)$. Where $m$ can be written as a linear combination of the actual curated differences $d_j$ of pixel values, $$\hat{m}(s) = \sum_{j=1}^{N}p_j(s) y_j$$ 	where $p(s)^{T} = (p_1(s),...,p_N(s))$ are the rows of the hat matrix specified by the local quadratic approximation. The estimated standard deviation of the local estimate $\hat{m}$ is $\hat{S}(s) = \hat{\sigma}\lVert (s) \rVert$. From these estimates, a T-type statistics is derived as,	$$T(s)  = \frac{\hat{m}(s)}{\hat{S}(s)}$$ $T(s)$ follows t-distribution and the null hypothesis is rejected at $s$ when $T(s) > t_{1-\alpha}(\sigma_1^2/\sigma_2)$ under a significance level of $\alpha$ with the degrees of freedom $n-6$, and $\sigma_1$ and $\sigma_2$  obtained by two-moment chi-square approximations \cite{WCleveland1988,JSun1994,XWang2005}. T-type statistics is different from the conventional T-statistics because it uses the weighted average of the values near $s$ whereas conventional T-statistics uses simple average. Therefore, $T(s)$ and $T(s')$ can be correlated for our T-type statistics whereas the test statistics are independent to each other  in the conventional T-tests.		
\subsection{FDR-controlled P-Map}		
The P-Map is generated from multiple comparison controlled p-values derived from the distribution of the T-type statistics. The most important problem in generating P-Map is overcoming the multiplicity problem for simultaneous comparisons.  Family-wise error rate (FWER) such as Bonferroni correction and false discovery rate(FDR) are the most common methods used to control for multiple comparisons. Section 4.2 of \textit{Zhang(2005)}\cite{ZZhang2005} discussed the relationship between FWER and FDR and we decided to use FDR for \pkg{LISA}. Specifically, we used the Benjamini and Hochberg's FDR-controlled  method \cite{YBenjamini1995}. The details of the method are introduced in section 4.2 Multiple testing problem of \textit{Wang et al.(2006)}\cite{XWang2006}. \pkg{LISA} provides `2D' and `3D' options in displaying the P-map.	
			
\section[Example]{Application}
In this section, an application of \pkg{LISA} on Neuromuscular Electrical Stimulation(NMES) experiment \cite{NBergstrom2004} is introduced. NMES is a treatment proposed to improve pressure distribution at the seating or lying support area for patients with mobile difficulties. Patients are at a risk of developing pressure ulcers when seated or lying down for a long time due to the blocking of the blood flow. NMES is intended to improve the blood flow for healthier regional tissue. In this application, the effect of NMES is evaluated by comparing the sequences of the pressure mat images before and after the treatment. A step by step process of using \pkg{LISA} will be introduced to conduct the comparison.\\
	
Generic \pkg{LISA} function : \\
	
\begin{lstlisting}
  lisa('filename1','filename2','scantype','nframe','nrow','ncol',
       'rowid','colid','preprocess','segtype','registtype','roi',
       'displaym','dimenpmap','parcomp')
\end{lstlisting}
	
\begin{longtable}{ l p{12cm} }
	\code{filename1} & Name of the csv file that contains the first sequence of images\\ 
	\code{filename2} & Name of the csv file that contains the second sequence of images\\
	\code{scantype} &  Data scanning method. Possible options include\\
	&\code{`row'} : using a row identifier (required arguments : nrow, rowid)\\
	&\code{`col'} : using a column identifier (required arguments : colid)\\
	&\code{`no'} : using a blank row\\
	\code{nframe} &Number of frames in each sequence\\
	\code{nrow} &Number of rows in one frame or the height of the image\\ 
	\code{ncol} &Number of columns in one frame or the width of the image\\
	\code{rowid} &A characters string that identifies the start of a frame\\
	\code{colid} &An integer number that identifies the column where the frame numbers are stored\\
	\code{preprocess} & Logical value to inform if the images have been pre-processed. Possible options include\\
	&\code{`T'} : True\\
	&\code{`F'} : False (default)\\
	\code{segtype} &Type of image segmentation. Possible options include\\
	&\code{`auto'} : De-noises the image by choosing a cutoff value to divide noise to the region of interest by using a mixture of normal distributions of the intensity values (default)\\
	&\code{`manual'} : De-noises the image through user-chosen cutoffs for the intensity values to divide noise to the region of interest\\
	\code{registtype} &Type of image registration. Possible options include\\
	&\code{`auto'} : Rotates and/or shifts images based on a mid-line found by a linear regression (no user interface is required) (default)\\
	&\code{`manual'} : Rotates and/or shifts images based on a line specified by two points selected manually by a user (the first click sets a reference point and the second click draws the reference line)\\
	\code{roi} &Select region of interest after image registration. Possible options include\\
	&\code{`T'} : True\\
	&\code{`F'} : False (default)\\
	\code{display} &Output option. Possible options include\\
	&\code{`basic'} : Only displays the P-Movie (movie of multiple comparison adjusted p-values) (default)\\
	&\code{`all'} : Displays O-Movies (movies of the original two sequences of images), R-Movies (movies of the registered sequence of images), D-Movie (movie of the curated difference), S-Movie (movie of the non-parametrically smoothed differences), T-Movie (movie of variance-adjusted smoothed differences), and P-Movie\\
	\code{dimenpmap} &Select the dimension of P-Movie. Possible options include\\
	&\code{2} : Two dimensions (default)\\
	&\code{3} : Three dimensions\\
	\code{parcomp} &Parallel computing. Possible options include\\
	&\code{`T'} : True\\
	&\code{`F'} : False (default)\\
\end{longtable}
	
\subsection{Read in Data}
The argument \code{`scantype'} takes one of the three possible values. When \code{scantype = `row'}, \pkg{LISA} finds the first row that matches \code{`rowid'}, then \code{`nrow'} rows after the \code{`rowid'} are stored as a frame. The search continues until the frame after the last \code{`rowid'} is read in. When \code{scantype = `col'}, \pkg{LISA} scans the csv file, and read in frames using the frame numbers found under the \code{`colid'} column. When \code{scantype = `no'}, \pkg{LISA} uses blank rows as row ids, and read in \code{`nrow'} rows after each blank row as individual frames. After all data are read in, \pkg{LISA} will have two arrays with \code{nframe} number of cells having \code{`nrow'} by \code{`ncol'} matrix of the intensity values in each cell.
		
\subsection{Image Processing}
A user can skip the image processing step if it's unnecessary by setting \code{`preprocess'} to \code{`T'}. Otherwise, a user can set \code{`preprocess'} to \code{`F'} to process the images as follows.
		
\subsection{Select Region of Interest}
\begin{figure}[h!]
\begin{minipage}{.5\linewidth}
	\centering
	\subfloat[][Select ROI for 1\textsuperscript{st} sequence]{\includegraphics[width=6cm]{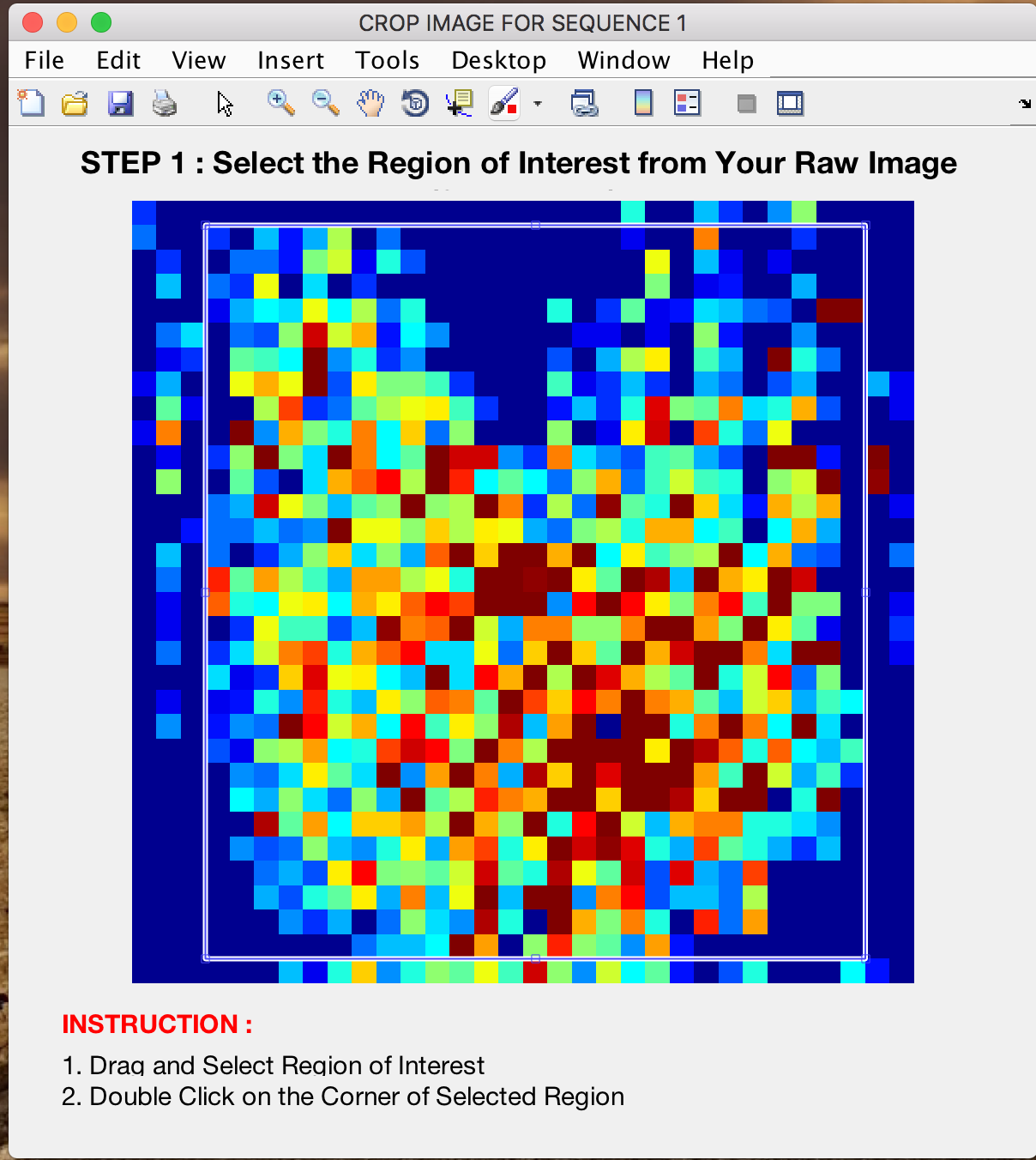}\label{<figure1>}}
\end{minipage}
\begin{minipage}{.5\linewidth}
	\subfloat[][Select ROI for 2\textsuperscript{nd} sequence]{\includegraphics[width=6cm]{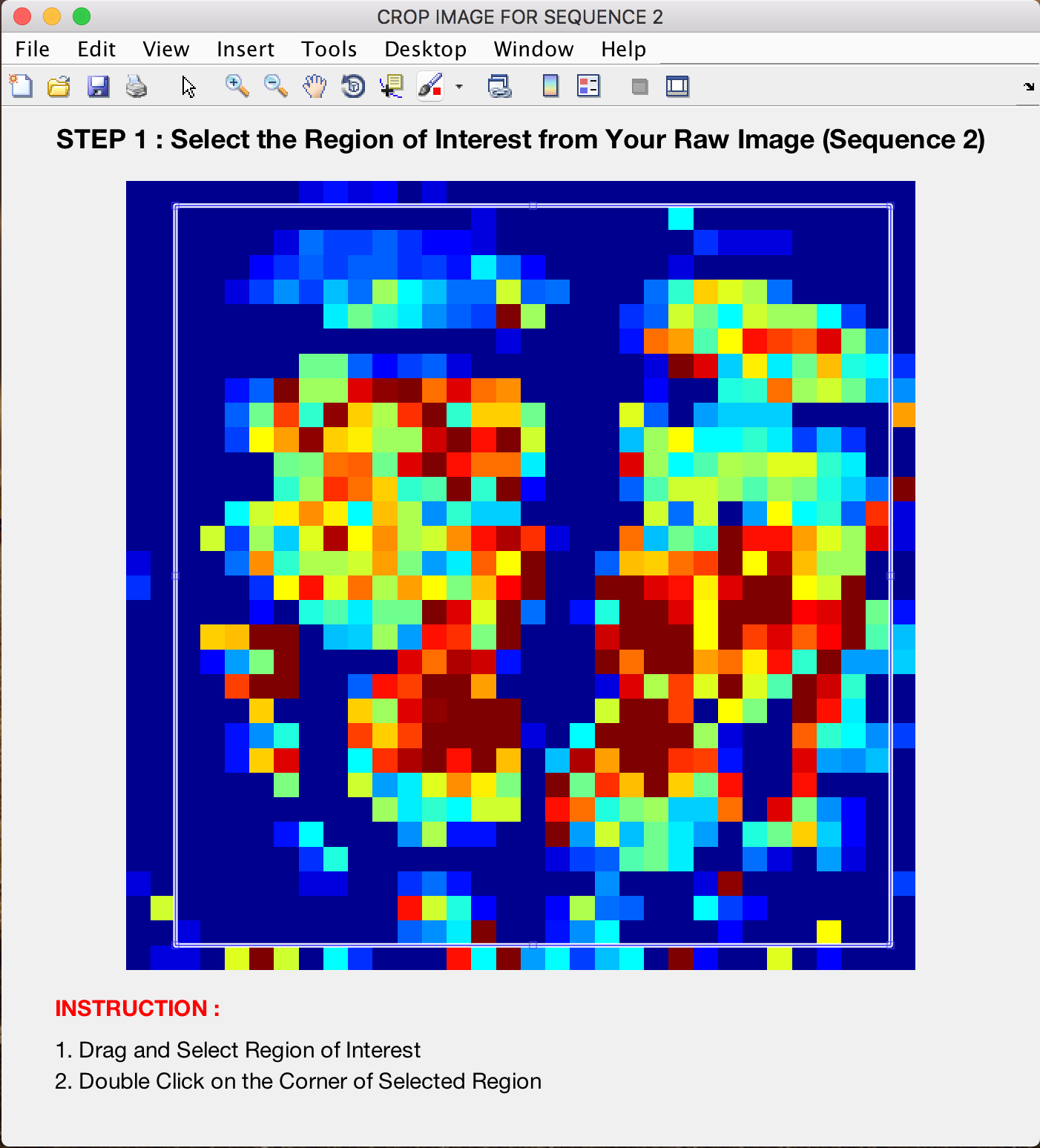}\label{<figure2>}}
\end{minipage}
\caption{Example of ROI selection}
\label{fig:1roi}
\end{figure}

\pkg{LISA} shows the first images of each sequence to select ROI as the first step of image processing. An example is displayed in Figure \ref{fig:1roi} and specific instructions on how to choose ROI are provided on the screen. A user needs to drag a square to roughly select the ROI. It is advised to choose a region large enough to enclose all ROI. Between the two regions selected from each sequence, the size of the larger region will be used to crop the images. 	
	
\subsection{Image Segmentation}
In the image segmentation step, noises to ROI are identified using the pixel intensity values. Automatic and manual image segmentation options are available in \pkg{LISA} as described in section 3.3. The following is an example. \\

(Automatic segmentation)
\begin{enumerate}
	\item {Select ROI : A user can use a multi-point polygon to select detailed ROI as shown in Figure \ref{fig:segroi}.}		
	\begin{figure}[h!]
		\begin{minipage}{.5\linewidth}
			\centering
			\subfloat[][Select detalied ROI for 1\textsuperscript{st} sequence]{\includegraphics[width=6cm]{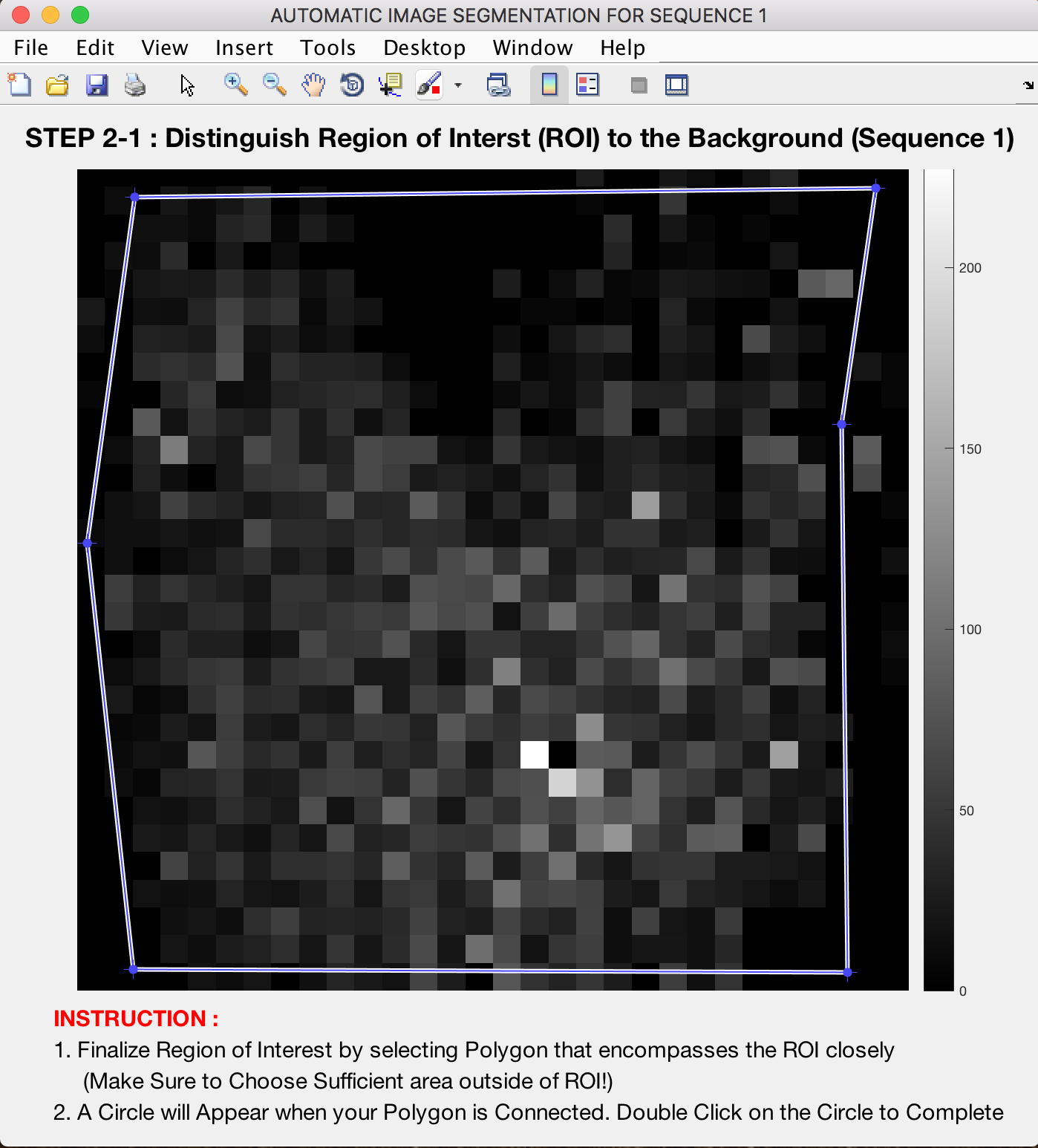}\label{<figure1>}}
		\end{minipage}
		\begin{minipage}{.5\linewidth}
			\subfloat[][Select detailed ROI for 2\textsuperscript{nd} sequence]{\includegraphics[width=6cm]{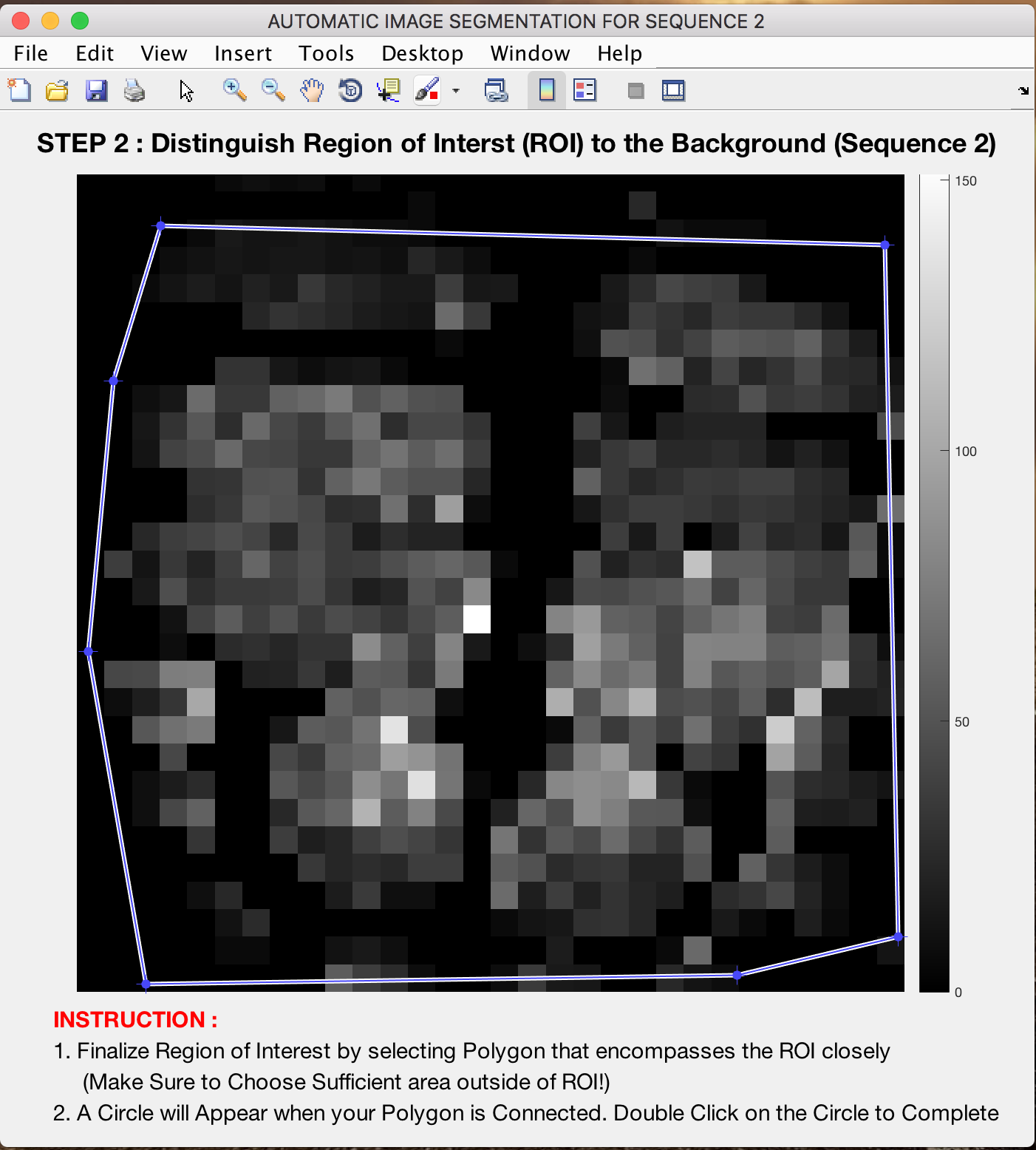}\label{<figure2>}}
		\end{minipage}
		\caption{Example of ROI selection for automatic image segmentation}
		\label{fig:segroi}
	\end{figure}		
	\item {Select groups for mixture of Gaussian distribution : A user can change the number of bins and define visually distinct groups from the distribution of intensity as shown in Figure \ref{fig:seggaussian}.}		
	\begin{figure}[h!]
		\begin{minipage}{.5\linewidth}
			\centering
			\subfloat[][Select groups for 1\textsuperscript{st} sequence]{\includegraphics[width=6cm]{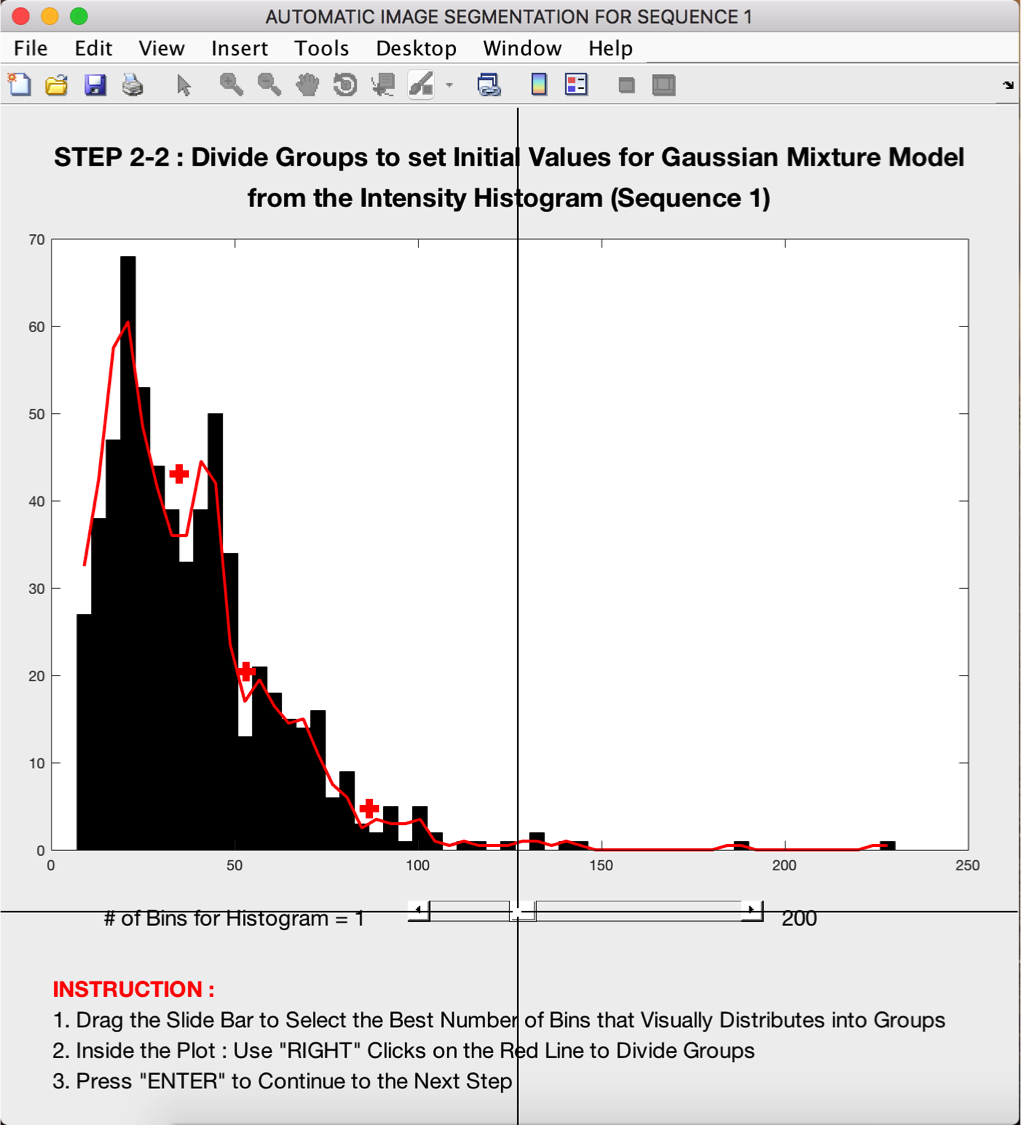}\label{<figure1>}}
		\end{minipage}
		\begin{minipage}{.5\linewidth}
			\subfloat[][Select groups for 2\textsuperscript{nd} sequence]{\includegraphics[width=6cm]{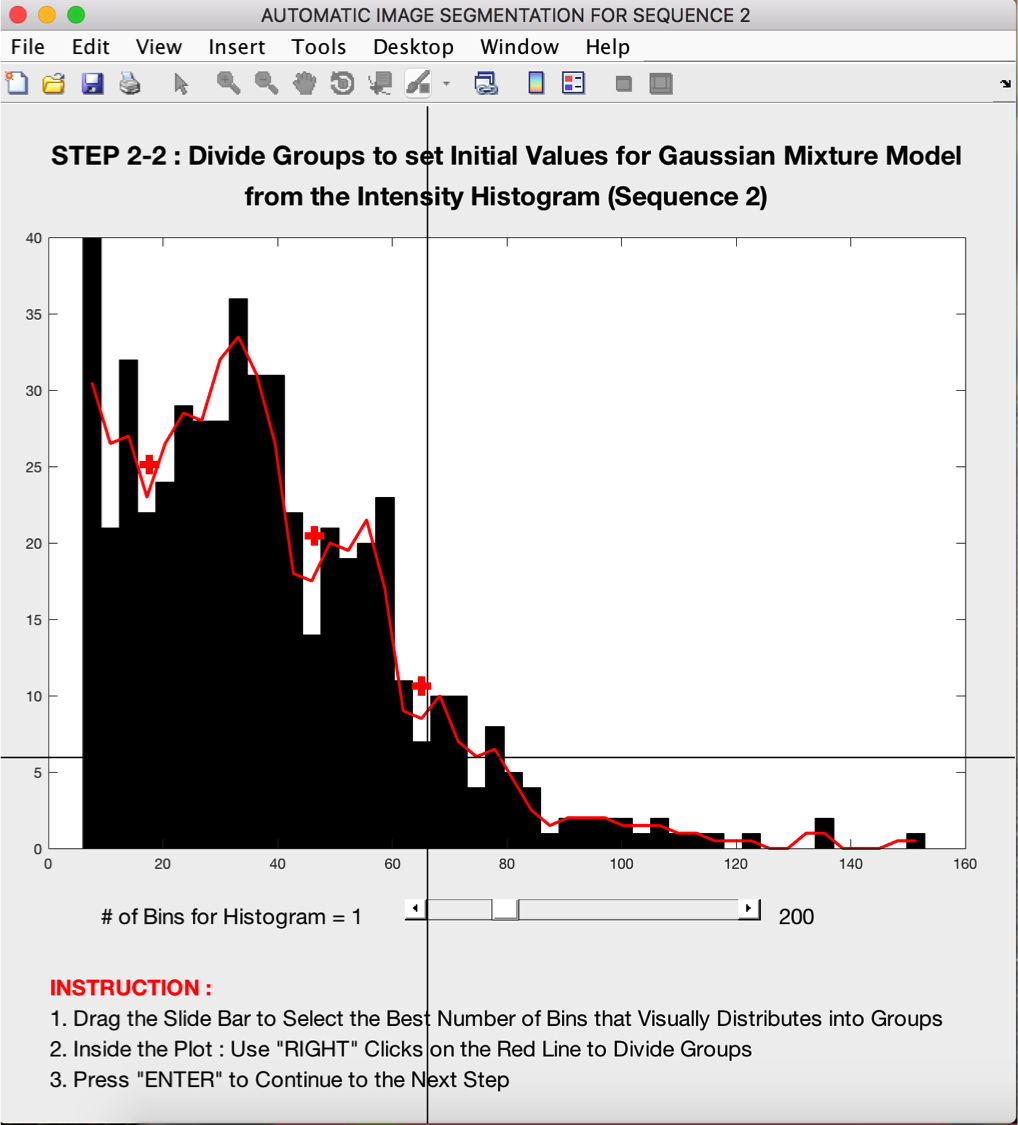}\label{<figure2>}}
		\end{minipage}
		\caption{Example of groups selection for automatic image segmentation}
		\label{fig:seggaussian}
	\end{figure}		
\end{enumerate}
		
(Manual segmentation)		
\begin{enumerate}
	\item {Pick cutoff value : A user should select one cutoff value from the histogram of the intensity values to distinguish the noise and ROI as shown in Figure \ref{fig:segcut}.}		
	\begin{figure}[h!]
		\begin{minipage}{.5\linewidth}
			\centering
			\subfloat[][Select cutoff value for 1\textsuperscript{st} sequence]{\includegraphics[width=6cm]{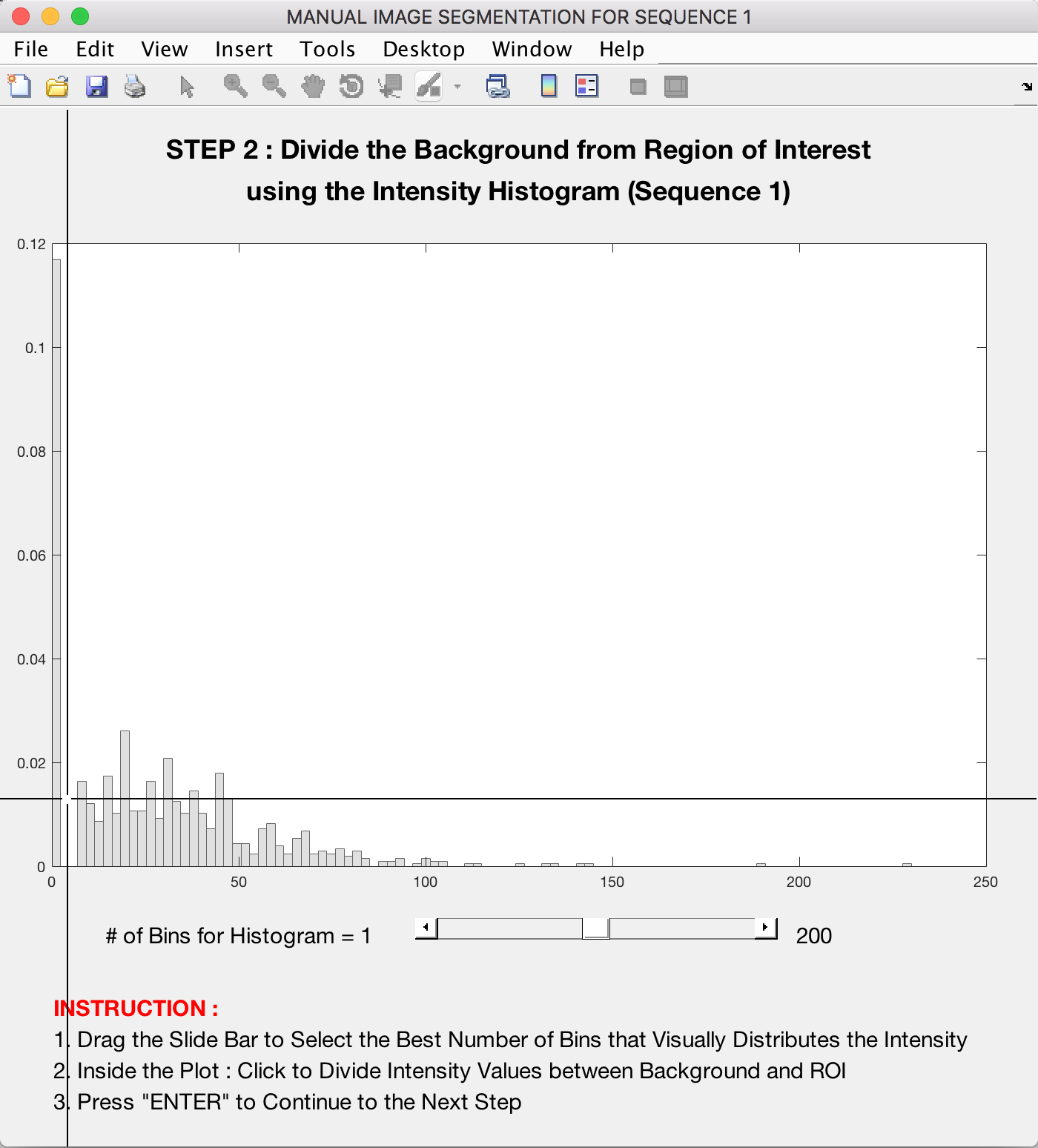}\label{<figure1>}}
		\end{minipage}
		\begin{minipage}{.5\linewidth}
			\subfloat[][Select cutoff value for 2\textsuperscript{nd} sequence]{\includegraphics[width=6cm]{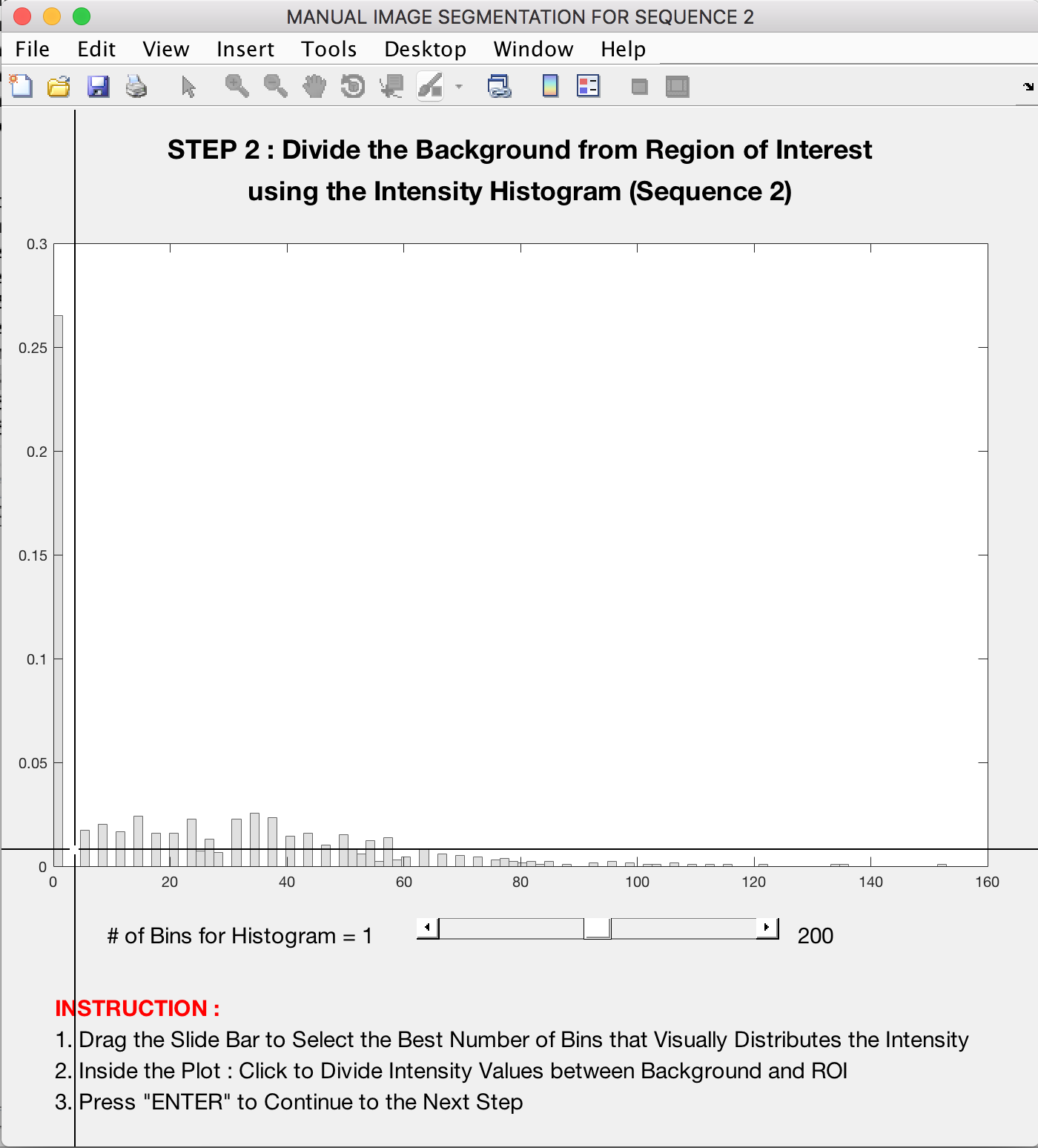}\label{<figure2>}}
		\end{minipage}
		\caption{Example of selecting cutoff value for manual image segmentation}
		\label{fig:segcut}
	\end{figure}		
\end{enumerate}	
			
\subsection{Image Registration}
\begin{figure}[h!]
	\begin{minipage}{.5\linewidth}
		\centering
		\subfloat[][Select reference point and line for 1\textsuperscript{st} sequence]{\includegraphics[width=6cm]{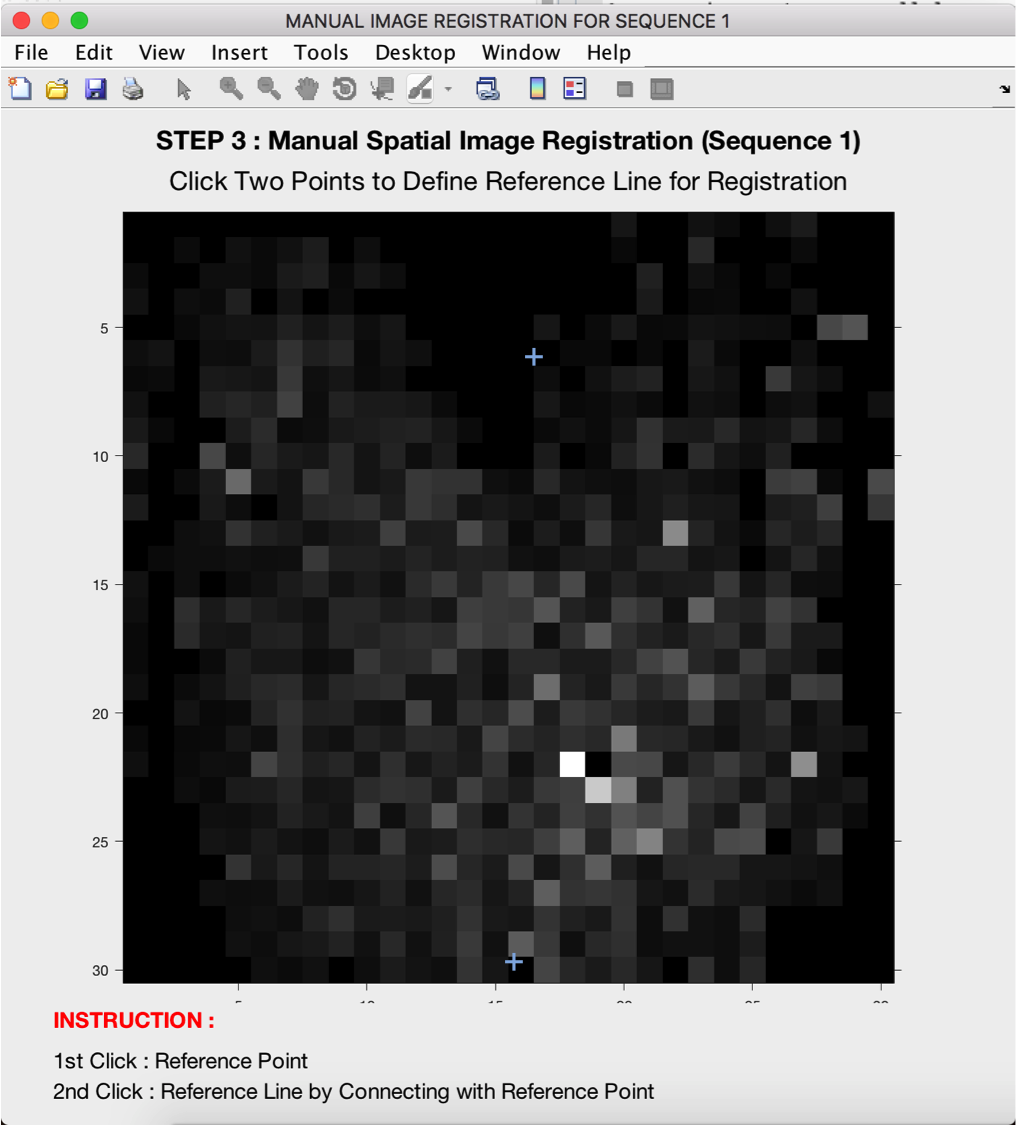}\label{<figure1>}}
	\end{minipage}
	\begin{minipage}{.5\linewidth}
		\subfloat[][Select reference point and line for 2\textsuperscript{nd} sequence]{\includegraphics[width=6cm]{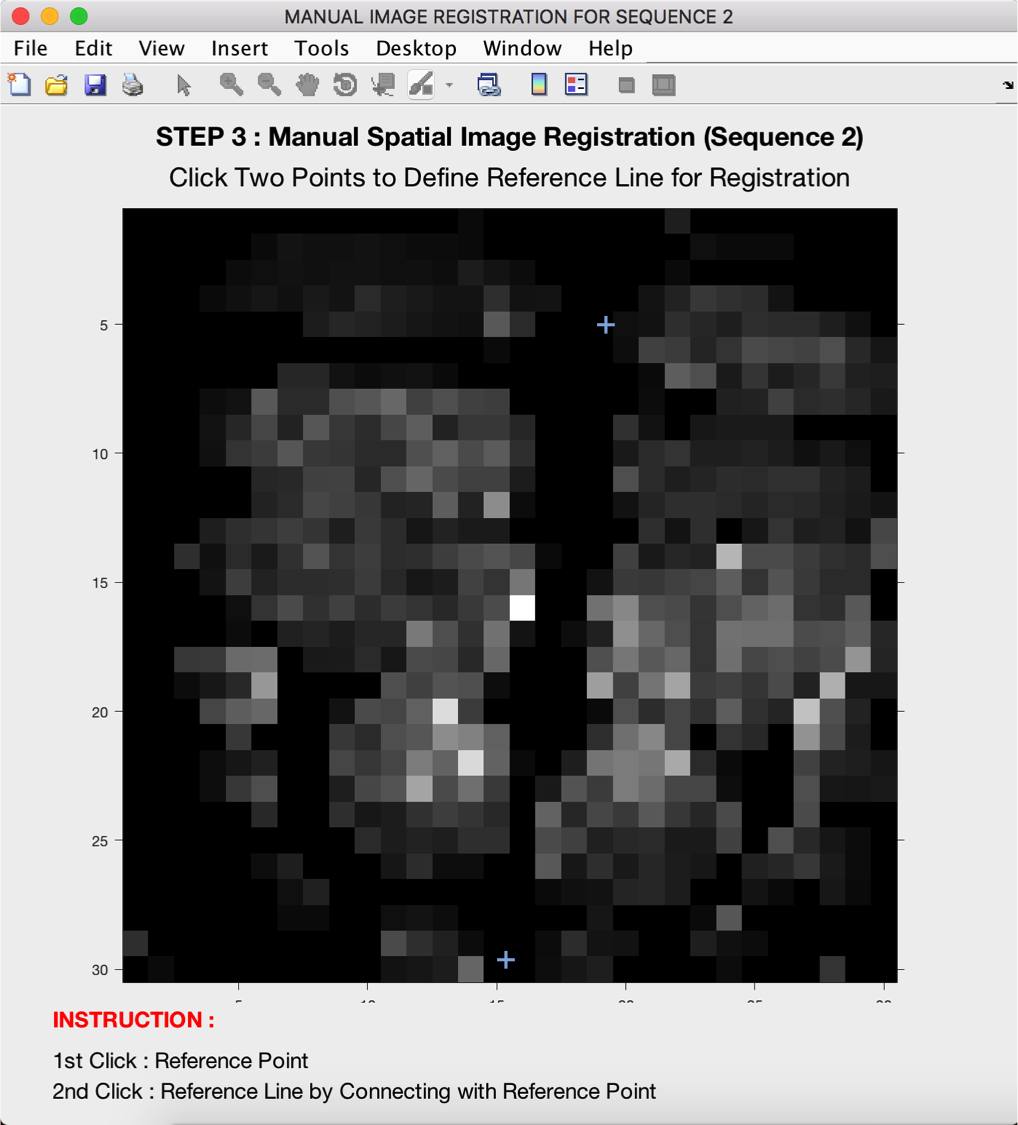}\label{<figure2>}}
	\end{minipage}
	\caption{Example of points selection for manual image registration}
	\label{fig:manureg}
\end{figure}

Image registration can be done automatically or manually as described in section 3.4. Automatic registration does not require any additional input from a user. Manual registration requires a user to select two points : a reference point and a second point to form a reference line using the first image of each sequence as shown in Figure \ref{fig:manureg}.
		
\subsection{Image Alignment Check}
\begin{figure}[h!]
	\centering
	\includegraphics[width=6cm]{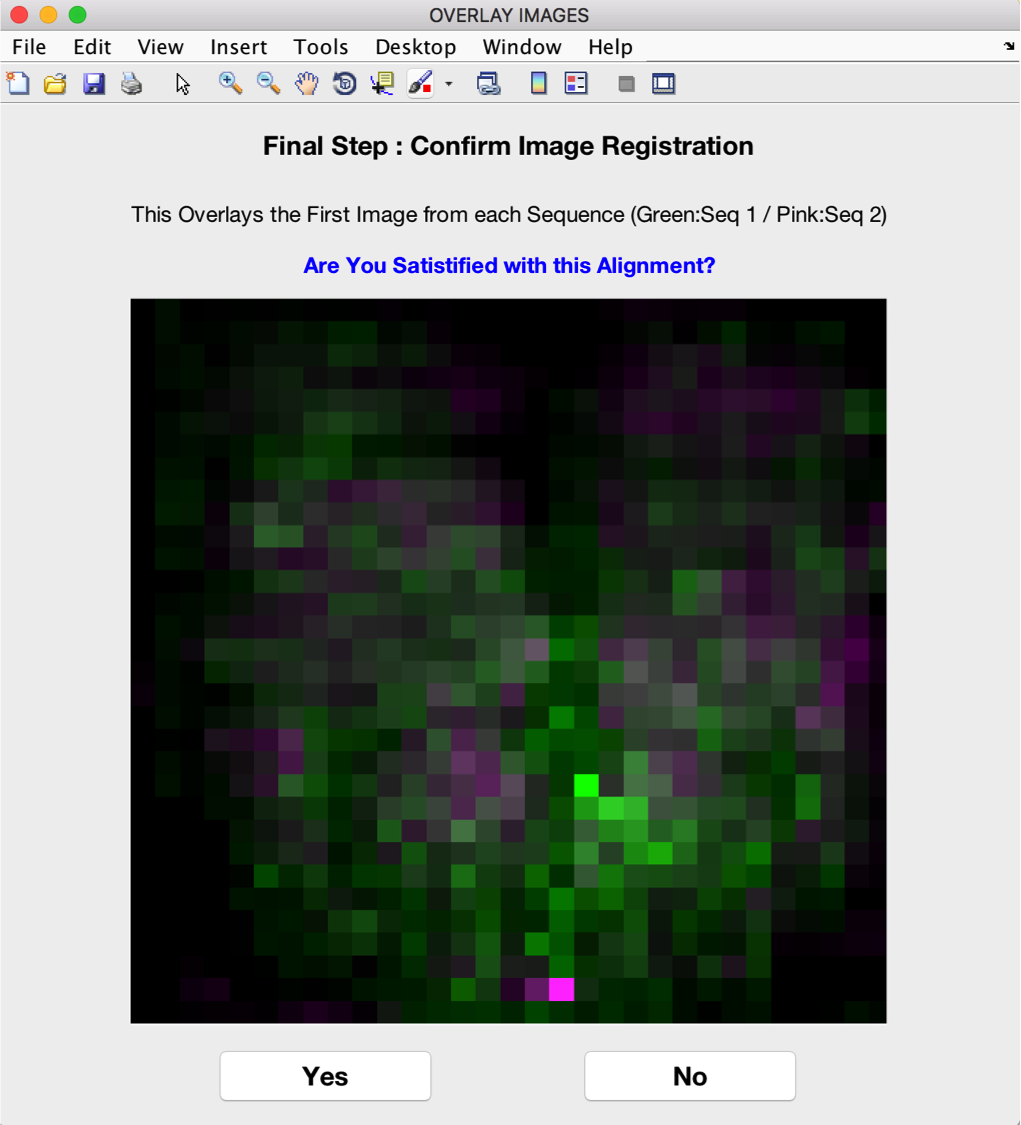}
	\caption{Checking image alignment before statistical analysis}
	\label{fig:overlap}
\end{figure}

Before conducting statistical comparison, \pkg{LISA} overlaps the first images of the two sequences which have been processed to confirm whether the image segmentation and image registration is satisfied by a user. If a user is satisfied and selects `Yes', then \pkg{LISA} continues to statistical analysis but if a user selects `No' \pkg{LISA} provides suggestions to the user on image processing in a pop-up window and terminates the process. An example of the overlapped image is shown in Figure \ref{fig:overlap}.
		
\subsection{Output Movies}
Depending on the option a user selects, \pkg{LISA} can provide 8 different movies :
\begin{enumerate}
	\item O-movie 1 (Figure 17-a) : Movie of the actual images from the first sequence
	\item O-movie 2 (Figure 17-b) : Movie of the actual images from the second sequence 
	\item R-movie 1 (Figure 17-c) : Movie of the registered images from the first sequence
	\item R-movie 2 (Figure 17-d) : Movie of the registered images from the second sequence 
	\item D-movie (Figure 17-e) : Movie of the curated difference 
	\item S-movie (Figure 17-f) : Movie of the non-parametrically smoothed difference
	\item T-movie (Figure 17-g) : Movie of the T-type statistics 
	\item P-movie (Figure 17-h) : Movie of FDR-controlled p-values 
	\end{enumerate}
		
\begin{figure}
	\centering
	\subfloat[O-movie 1]{\includegraphics[width = 3.5cm]{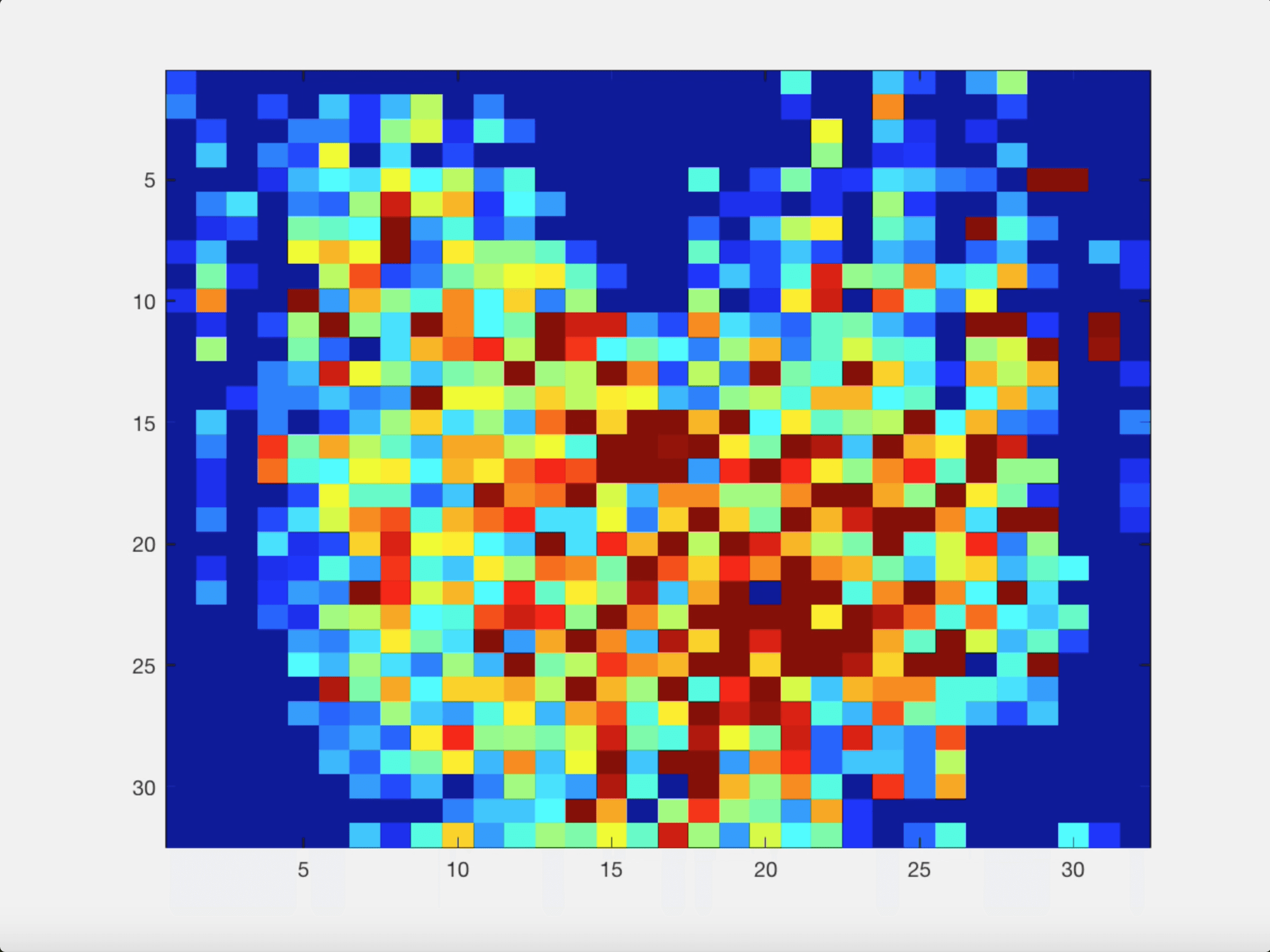}} 
	\subfloat[O-movie 2]{\includegraphics[width = 3.5cm]{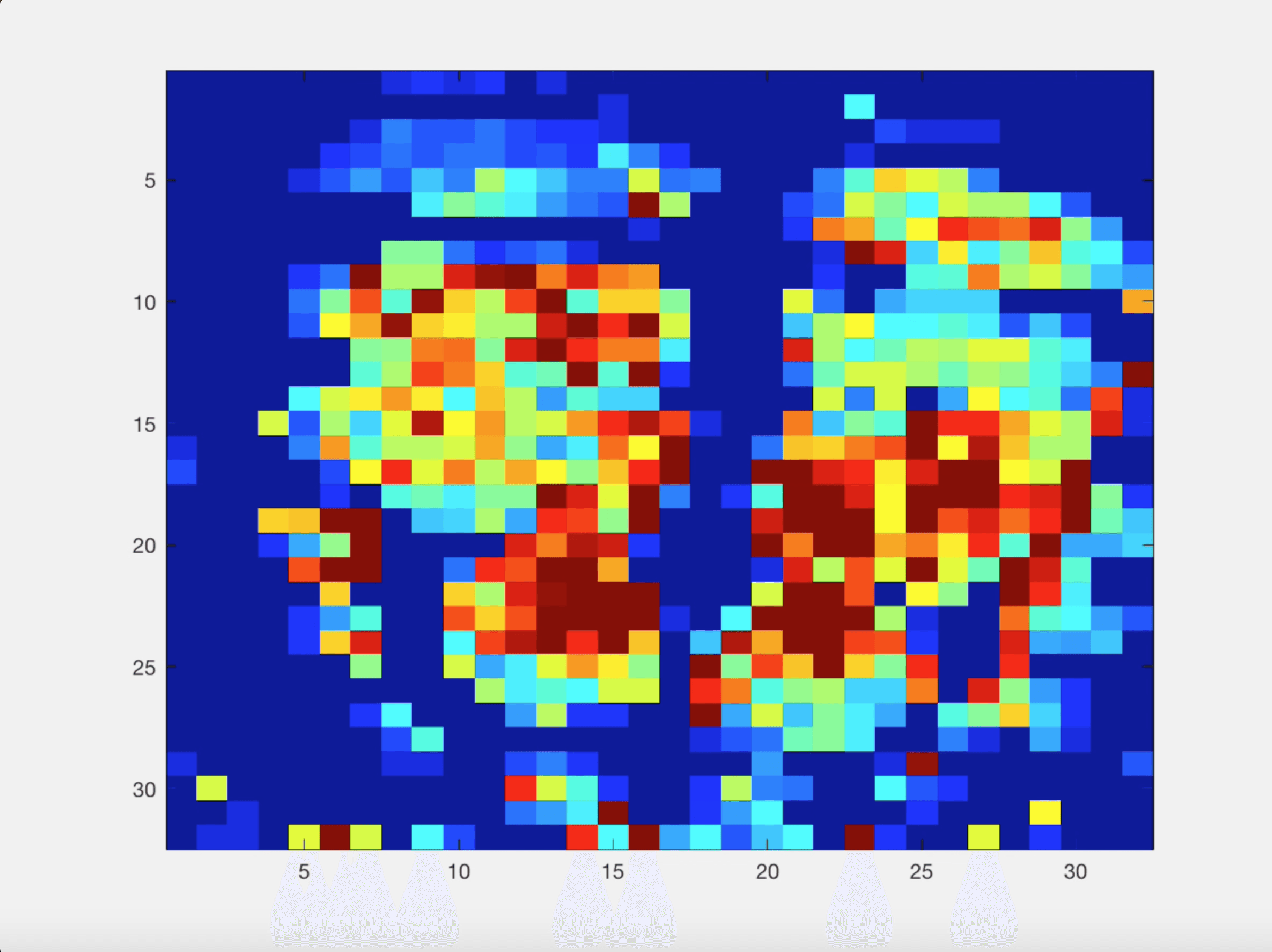}}    
	\subfloat[R-movie 1]{\includegraphics[width = 3.5cm]{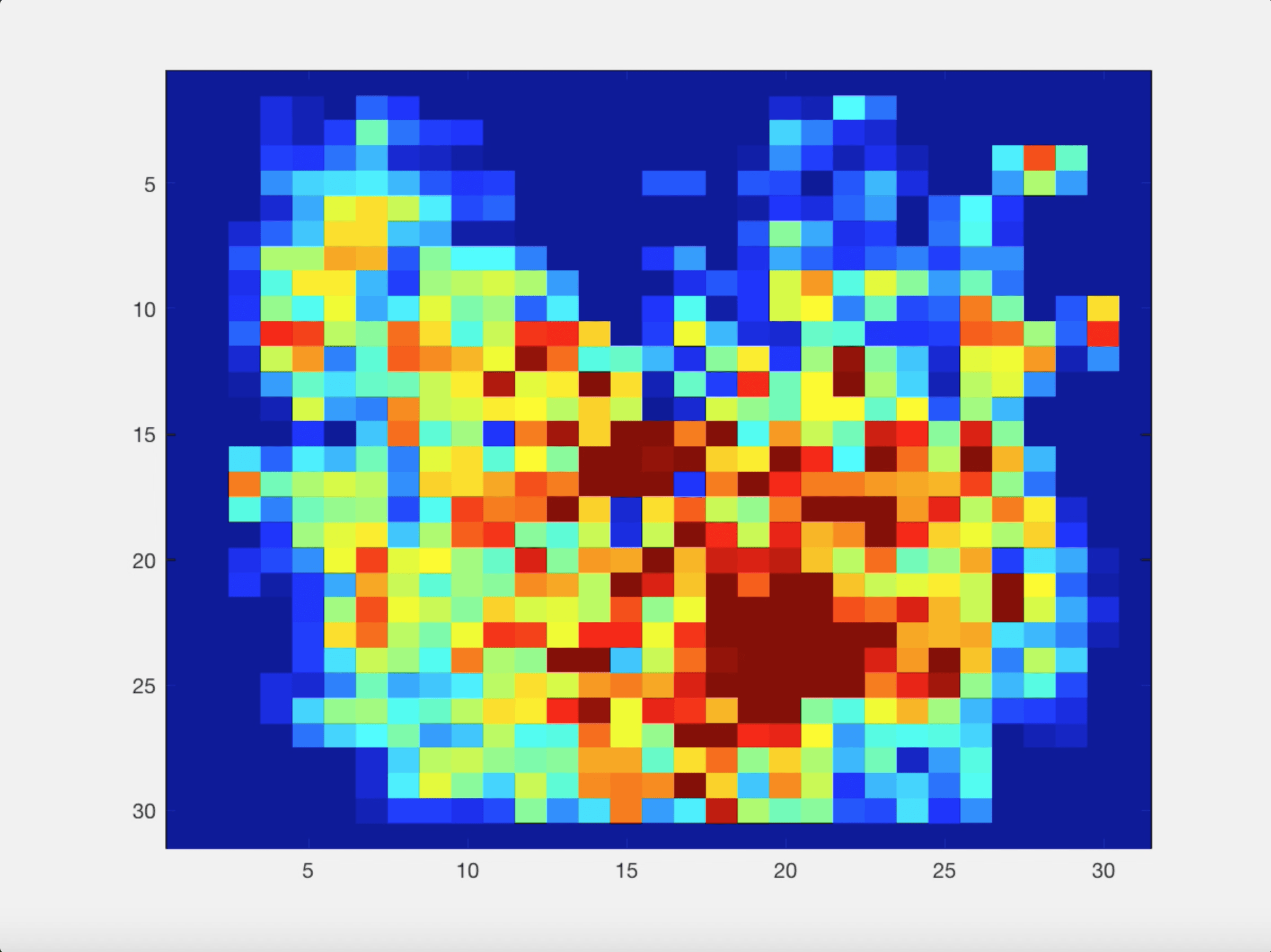}}
	\subfloat[R-movie 2]{\includegraphics[width = 3.5cm]{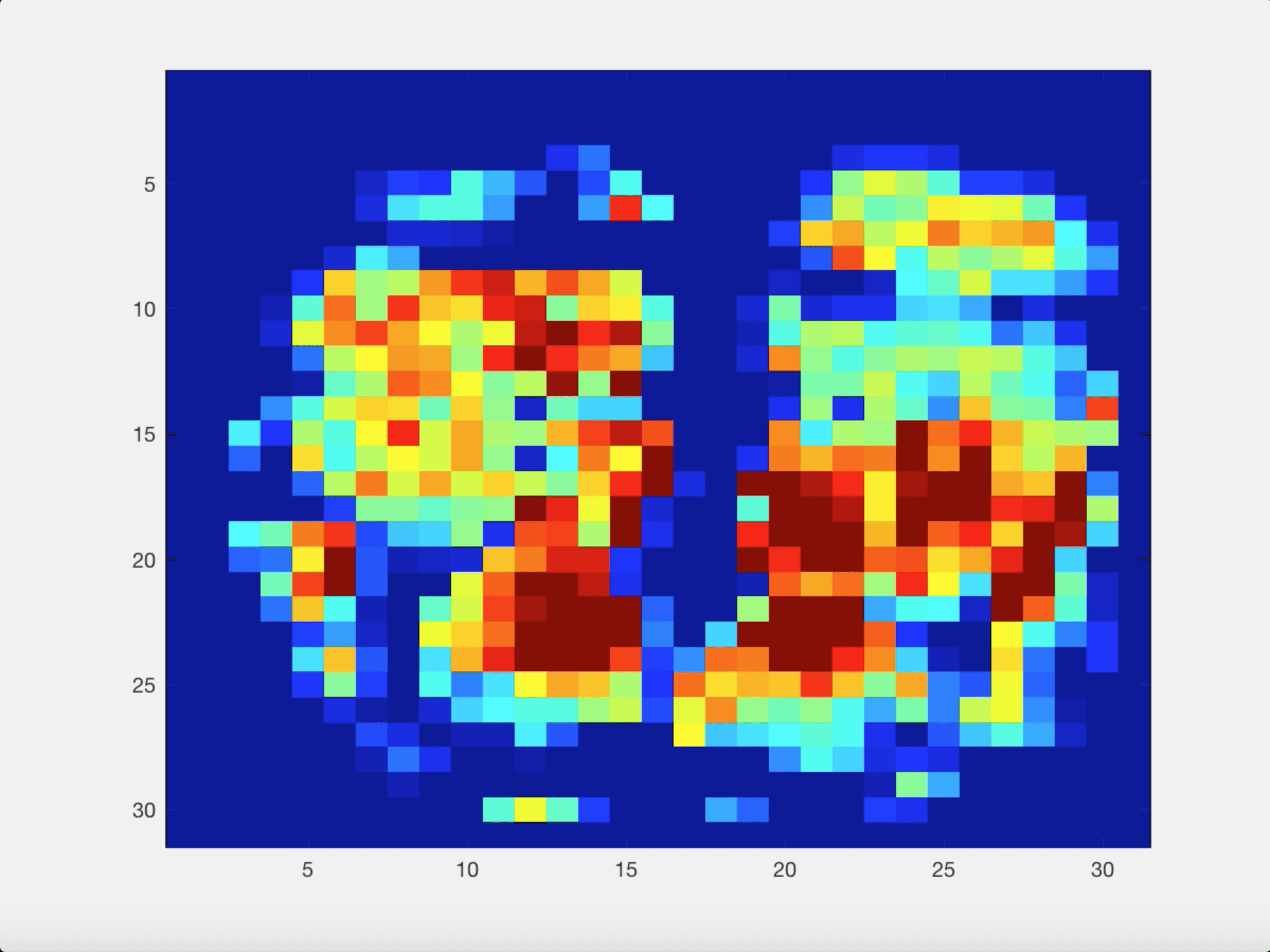}}\\ 
	\subfloat[D-movie]{\includegraphics[width = 3.5cm]{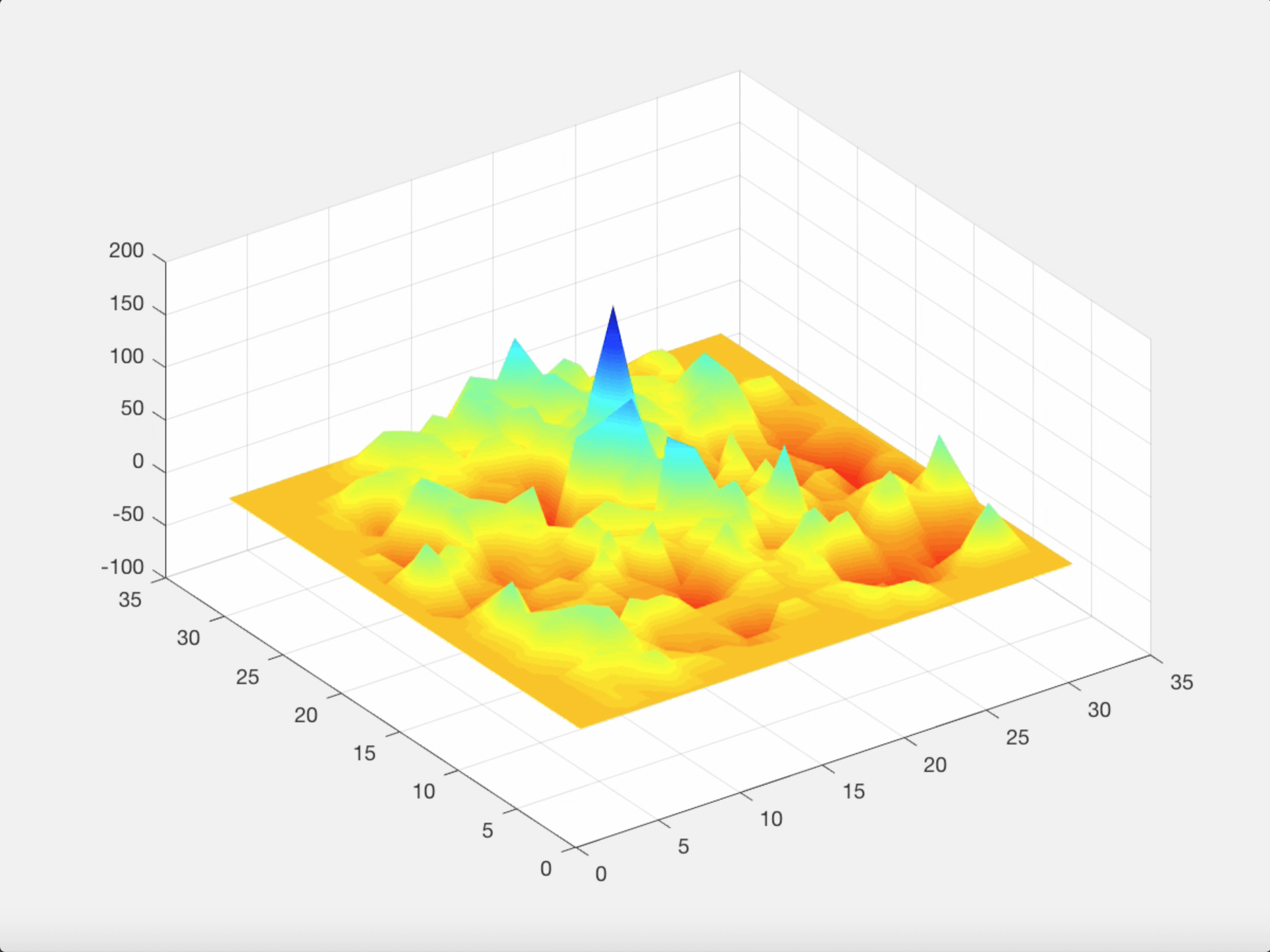}}
	\subfloat[S-movie]{\includegraphics[width = 3.5cm]{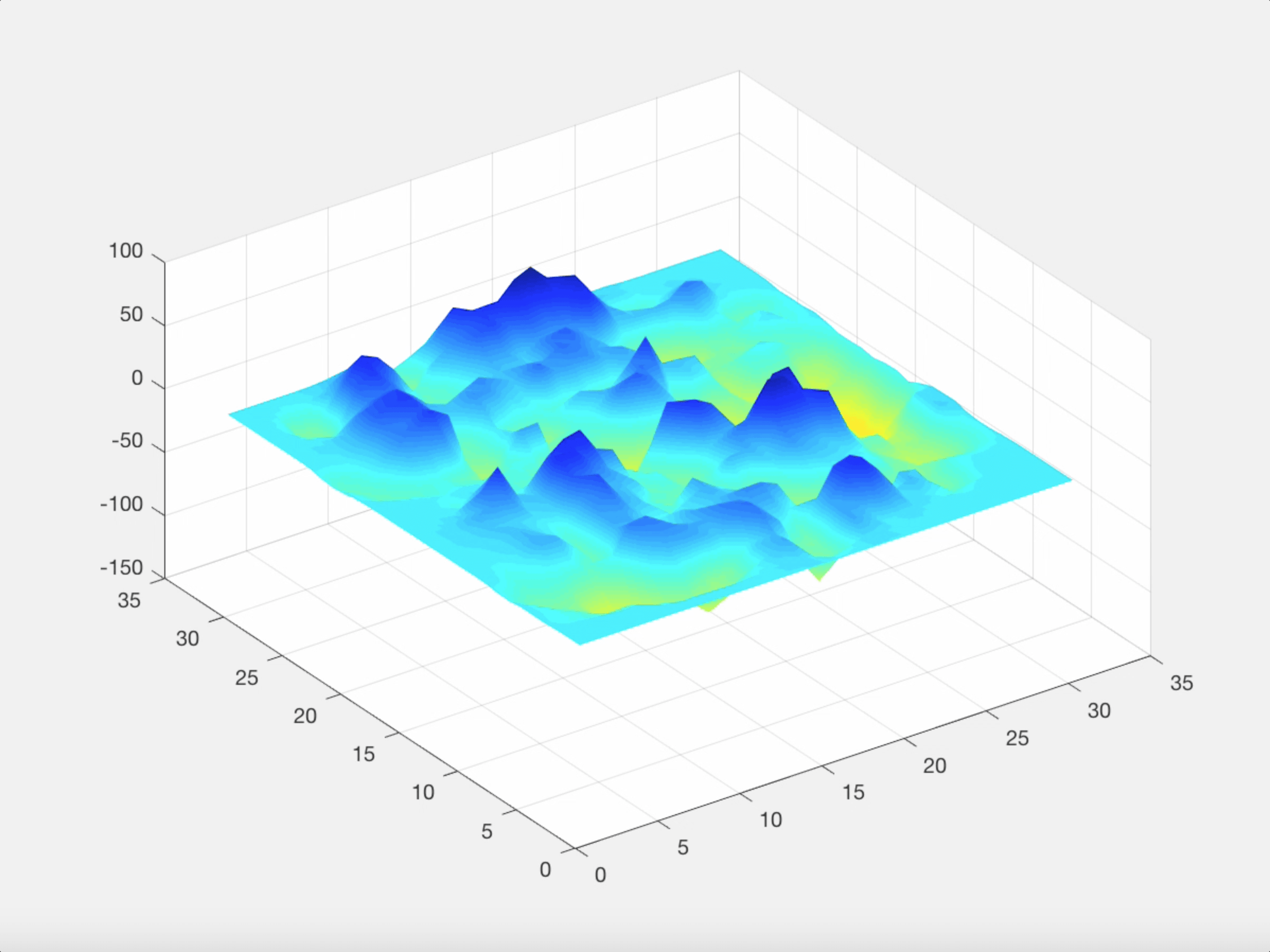}}
	\subfloat[T-movie]{\includegraphics[width = 3.5cm]{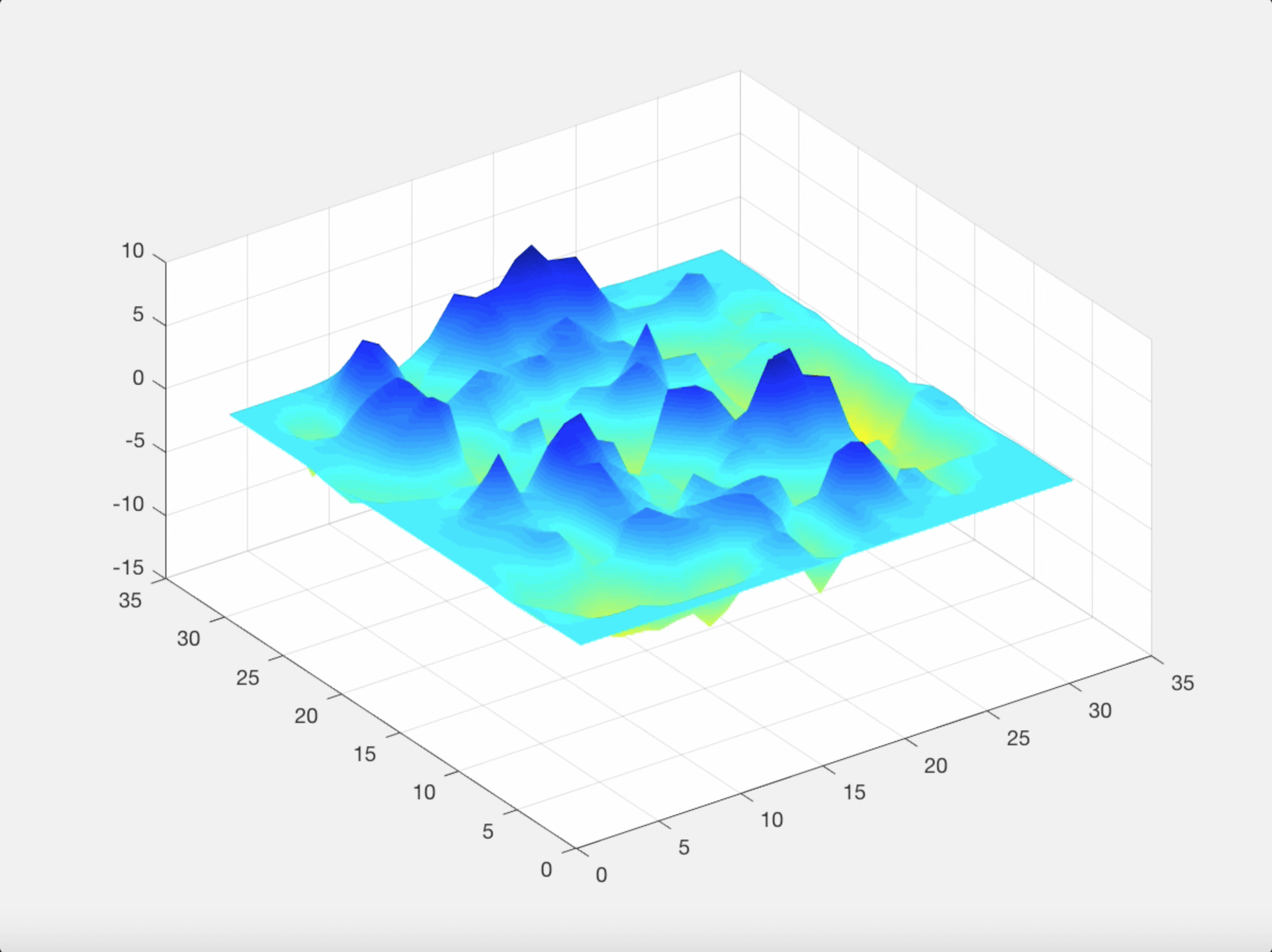}}
	\subfloat[P-movie]{\includegraphics[width = 3.5cm]{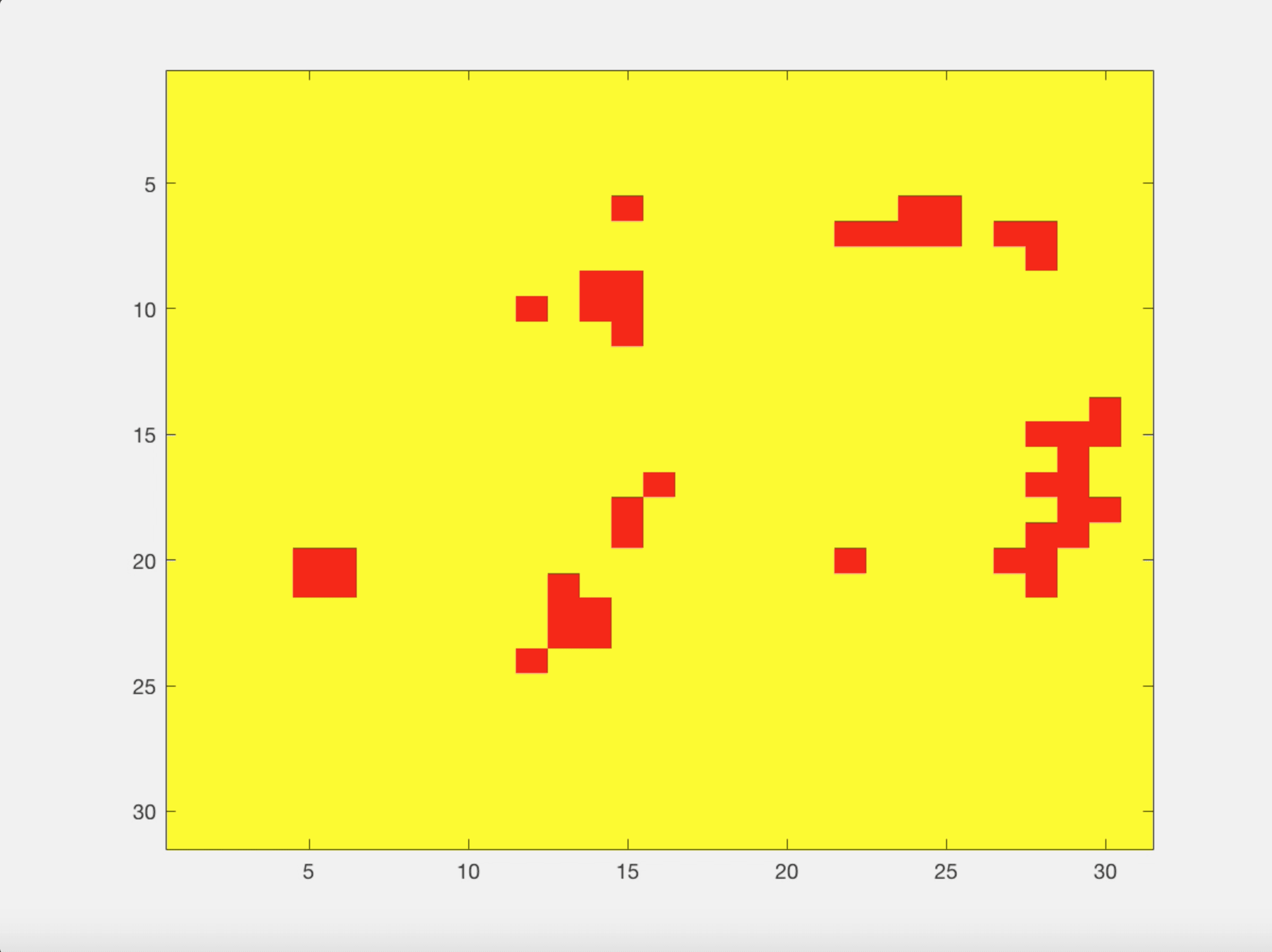}}
	\caption{Output movies}
	\label{fig:results}
\end{figure}

When \code{`basic'} is selected only P-map will be generated but when \code{`all'} is selected, all 8 movies will be generated for display. In D-movie, S-movie and T-movie, a blue peak signifies positive differences which indicate higher pressure in the first sequence compared to the second sequence, whereas, a red peak signifies the opposite. In P-movie, red areas are statistically significantly activated/different areas. The snapshots of the movies resulting from our data application using the seated pressure mats are introduced in Figure \ref{fig:results}.\\
		
\section[Discussion]{Discussion}
\pkg{LISA} is an image analysis package that processes and analyzes two sequences of images to find significant activation regions between them. In \pkg{LISA}, the essential components of image processing functionality include ROI selection, image segmentation and image registration and that of statistical analysis functionality include non-parametric mapping, smoothing and multiple comparison adjusted testing.

The current version of \pkg{LISA} was developed for and focused on sequence of images from seated pressure mats but the package can be extended to analyze other types of time series images such as satellite images, hyper-spectral chemical images, longitudinal magnetic resonance images. The interactive function during the image processing step makes it easier for users to use the package and provides users with visual evidence to derive more accurate image alignment and statistical comparison results. Statistical analysis is designed to reduce possible errors resulting from edge effects and to account for multiple comparisons using false discovery rate which makes the final result conservative and accurate. \pkg{LISA} also has a parallel computing function to help with computing issues for big data. High-resolution or long image sequences can be analyzed with the parallel computing function in a multi-core processing environment.
	
Depending on a user's needs and preference, other image processing tools can enhance the overall analysis experience, such as the Image Processing Toolbox(IPT) within \proglang{MATLAB}. Among the many options in IPT, a function such as multi-point registration could help users to derive more accurate image alignments depending on the characteristics of the image. The underlying method is identical to manual registration in \pkg{LISA}, but users can select multi-reference points for image translation. For details, please refer to the MathWorks website \cite{matimageregist}. Any other tools can be used for image processing as long as the images are well aligned and the results are saved in `csv' format to be consumed by \pkg{LISA}.
	
Development of a standalone \proglang{MATLAB} package \pkg{LISA} as an ensemble of elements from the LASR algorithm \cite{XWang2006} with additional features for image processing and graphical user interface is the start of our project in building an image comparison analysis package for large dynamic sequence of high-resolution images. The current version can handle relatively simple and static sequence of images. However, we aim to develop a next generation of \pkg{LISA} to be able to handle more complex and dynamic images, such as the entire body of a human and comparing images at many different time points. If we take a pressure map of a entire body of a human as an example, our future work will focus on segmented image processing and relevant statistical comparisons. This will allow us to compare and find significant activation regions independently separating by upper body, lower body, arms, legs and for any specific areas.	

Another area of our future development includes reducing the computing time. Even with the parallel computing option, the computing time for \pkg{LISA} can be relatively long depending on the size of the image and sequence and the number of cores in the computer because the image comparison is done pixel-by-pixel. Pixel-by-pixel comparison is the most comprehensive and exhaustive way to compare images but can also provide most accurate results relative to using measures. Therefore, we will be aiming to develop methods such as to distribute the statistical calculation to reduce the computing time working under the essential framework of pixel-by-pixel comparison.
			
\newpage
\bibliographystyle{unsrt} 
\bibliography{lisabib.bib}
\end{document}